\documentclass[pre,twocolumn,aps]{revtex4}
\usepackage{amsmath, amssymb}
\usepackage{graphicx}
\usepackage{hyperref}
\usepackage{subcaption}
\usepackage{caption}
\usepackage{xcolor}
\usepackage{float}
\usepackage{placeins}
\usepackage{tikz}
\begin{document}

\title{Topological Anderson insulator phases in one dimensional quasi-periodic mechanical SSH chains}
\author{Sayan Sircar}
\affiliation{Tata Institute of Fundamental Research, Hyderabad 500046, India}
\date{July 29, 2024}
\begin{abstract}
In this paper, we investigate the transition between topological phases in a Su-Schrieffer-Heeger (SSH) model composed of springs and masses in which the intracellular Aubry-André disorder modulates the spring constants. We analytically compute the eigenvectors and eigenvalues of the dynamical matrix for both periodic and fixed boundary conditions, and compare them with the dispersion spectrum of the original tight-binding SSH model. We observe the presence of a topological Anderson insulating (TAI) phase within a specific range of quasi-periodic modulation strength and calculate the phase transition boundary analytically. We examine the localization properties of normal modes using their inverse participation ratio (IPR) of eigenstates of the dynamical matrix, and the corresponding fractal dimension associated with quasiperiodic modulation. We also examine the stability of the TAI phase across a range of modulation strengths and comments on the presence of mobility edge that separate localized modes from non-localized ones. We demonstrate the fact that special analytical techniques are needed to compute an exact expression for mobility edges in such scenarios.
\end{abstract}

\maketitle

\section{INTRODUCTION}
The study of topological phases of matter \cite{PhysRevLett.49.405} has gardened significant research interest in the last decades. Among this, topological insulators \cite{RevModPhys.82.3045,RevModPhys.83.1057} is a fascinating phase characterized by bulk states with a gap at the Fermi level, while hosting conducting edge modes in the bulk gap, \cite{RevModPhys.90.015001} characterized by a non-trivial winding number, as per bulk-edge correspondence \cite{Prodan_2016}. The Su–Schrieffer–Heeger (SSH) model \cite{PhysRevLett.42.1698} is a fundamental system consisting of two bands and showing non-trivial band topology , has been introduced to examine conductivity and transport properties of polyacetylene chain, particularly focusing on fractional charge soliton excitations, \cite{RevModPhys.60.781,PhysRevD.13.3398} and non-trivial conducting edge modes \cite{PhysRevLett.110.180403} in open boundary condition (OBC). Conducting edge states are maintained by internal symmetries and are resilient to specific types of disorder. Disorder in translational invariant systems disrupts the translational invariance of the system, leading to the assumption that it could undermine the topological properties.  The emergence of topological Anderson insulators (TAIs) indicates that disorder can trigger the transition from topologically trivial to non-trivial phases, \cite{PhysRevLett.102.136806,PhysRevB.80.165316,PhysRevLett.103.196805,PhysRevB.89.224203,PhysRevLett.113.046802}.  Anderson localization was proposed in a model of disordered
 system \cite{PhysRev.109.1492, RevModPhys.57.287} first  in 1958, which since then has been performed in various experimental settings \cite{PhysRevLett.85.2360}.
 Typically, real materials exhibit Anderson-type disorder \cite{PhysRev.109.1492}, characterized by random potentials generated by limited concentration of contaminants dispersed through out the material. Depending on spatial dimensions, potential intensity $\delta$ has a critical value for example in three dimensions \cite{PhysRevLett.42.673}. In contrast, in one dimension, for an infinitely large system, most of the states remain localized for arbitrary finite value of $\delta$. \par
The emergence of twisted bilayer-graphene \cite{Gonçalves_2022} introduces a novel form of disorder known as quasi-periodic disorder. The best known model to track quasi-periodicity in condensed matter systems is Aubry-Andre-Harper (AAH) model. The 1D Aubry-Andre-Harper (AAH) model \cite{aubry1980analyticity} illustrates a metal-insulator transition governed by self-duality point \cite{Domínguez-Castro_2019}. This transition occurs when the quasi-periodic modulation strength surpasses a critical threshold. Aubry and Andre introduced a one dimensional hopping model in tight-binding approximation, in which electrons experience a sinusoidal electrostatic potential that is incommensurate with the lattice. They demonstrated that the transition occurs at $\delta^{AA}=2J$, where $J$ represents the nearest-neighbour electron hopping amplitude. In this scenario, all eigenmodes become exponentially localized for $\delta^{AA}>2J$ or extend as plane waves for $\delta^{AA}<2J$. These results differ from Anderson localization phenomenon that occurs in one dimensional systems. These conclusions have also been extended to non-Hermitian systems \cite{PhysRevLett.122.237601,PhysRevB.100.125157}. The presence of mobility edges in disordered systems, particularly in 3D Anderson insulators, has been demonstrated even in the 1D generalized AAH model. Compact analytic results for mobility edges under broken self-duality have been obtained \cite{PhysRevLett.104.070601,PhysRevB.101.174205,PhysRevLett.125.196604,PhysRevResearch.5.023044} and for slowly varying modulation incommensurate with the lattice \cite{PhysRevLett.61.2144,PhysRevB.41.5544,liu2018topological}.These works focus on calculating mobility edges that differentiate localized from non-localized modes. Recent studies have expanded this concept to anomalous mobility edges, which separate localized states from critical ones in one-dimensional disordered systems, differing from conventional mobility edges, \cite{liu2022anomalous}. Disorder-induced topological Anderson insulators (TAI) have been experimentally observed in 2D photonics \cite{article}, 1D wires with controllable random disorder \cite{Meier_2018} protected by chiral symmetry, superconductors with spin-orbit coupling \cite{unknown,article}. The TAI phase also emerges in this model (SSH) modulated with quasi-periodic intercell hopping, where most of the eigenstates are localized \cite{longhi2020topological,sircar2024disorderdriventopologicalphase}. Recently there are interest in exploring similar phases in several acoustic systems and elastic \cite{PhysRevLett.114.114301,PhysRevB.96.134307}. Hence its a natural urge to search for a 
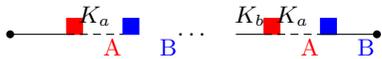
\begin{figure}[ht!]
\centering
\begin{tikzpicture}[scale=0.75]

\coordinate (F1) at (0,0);
\coordinate (A1) at (1,0);
\coordinate (B1) at (2,0);

\coordinate (A_end) at (4.5,0);
\coordinate (B_end) at (5.5,0);
\coordinate (F2) at (6.5,0);

\fill (F1) circle (2pt);
\fill (F2) circle (2pt);

\draw (F1) -- (A1);
\filldraw[red] (A1) rectangle ++(8pt, 8pt) node[below right=0.5em] {A};
\draw[densely dashed] (A1) -- (B1) node[midway, above] {$K_{a}$};
\filldraw[blue] (B1) rectangle ++(8pt, 8pt) node[below right=0.5em] {B};

\node at (3.25,0) {\ldots};

\coordinate (A_ellipsis) at (4,0);
\draw (A_ellipsis) -- (A_end) node[midway, above] {$K_{b}$};
\filldraw[red] (A_end) rectangle ++(8pt, 8pt) node[below right=0.5em] {A};
\draw[densely dashed] (A_end) -- (B_end) node[midway, above] {$K_{a}$};
\filldraw[blue] (B_end) rectangle ++(8pt, 8pt) node[below right=0.5em] {B};
\draw (B_end) -- (F2);

\end{tikzpicture}
\caption{Schematic representation of a mechanical SSH chain, where dashed lines represent intra-cellular springs ($K_{a}$) and solid lines denote inter-cellular springs ($K_{b}$). Each unit cell consists of $A$ and $B$ masses, with the black solid circles at the ends representing fixed points.}
\label{fig:QSW}
\end{figure}
similar phase in various mechanical settings \cite{PhysRevResearch.3.033012} as well. \par
In this article, starting from the model as shown in Ref.~\cite{lu2022exact}, we apply quasi-periodic modulation in on-site intra-cellular spring stiffness in a mechanical SSH chain, and calculate the topological phase transition boundary, localization properties with modulation $\delta^{AA}\cos(2\pi \beta j^{u}+\phi)$, where $\delta^{AA}$ is the quasi-periodic modulation strength and $u=1$.

\section{Mechanical spring-mass model}
This section provides introduction to the SSH mechanical model used, and examines its topological properties in real space, akin to Ref.~\cite{sircar2024disorderdriventopologicalphase}.

\subsection{Dynamical matrix of the spring-mass chain}
We are studying a one-dimensional finite chain whose ends are held fixed (FBC), and is made up of N unit cells \cite{PhysRevResearch.3.033012,PhysRevB.109.195427,sircar2024disorderdriventopologicalphase}. 
\begin{figure}[ht!]
    \centering
    \includegraphics[width=\linewidth]{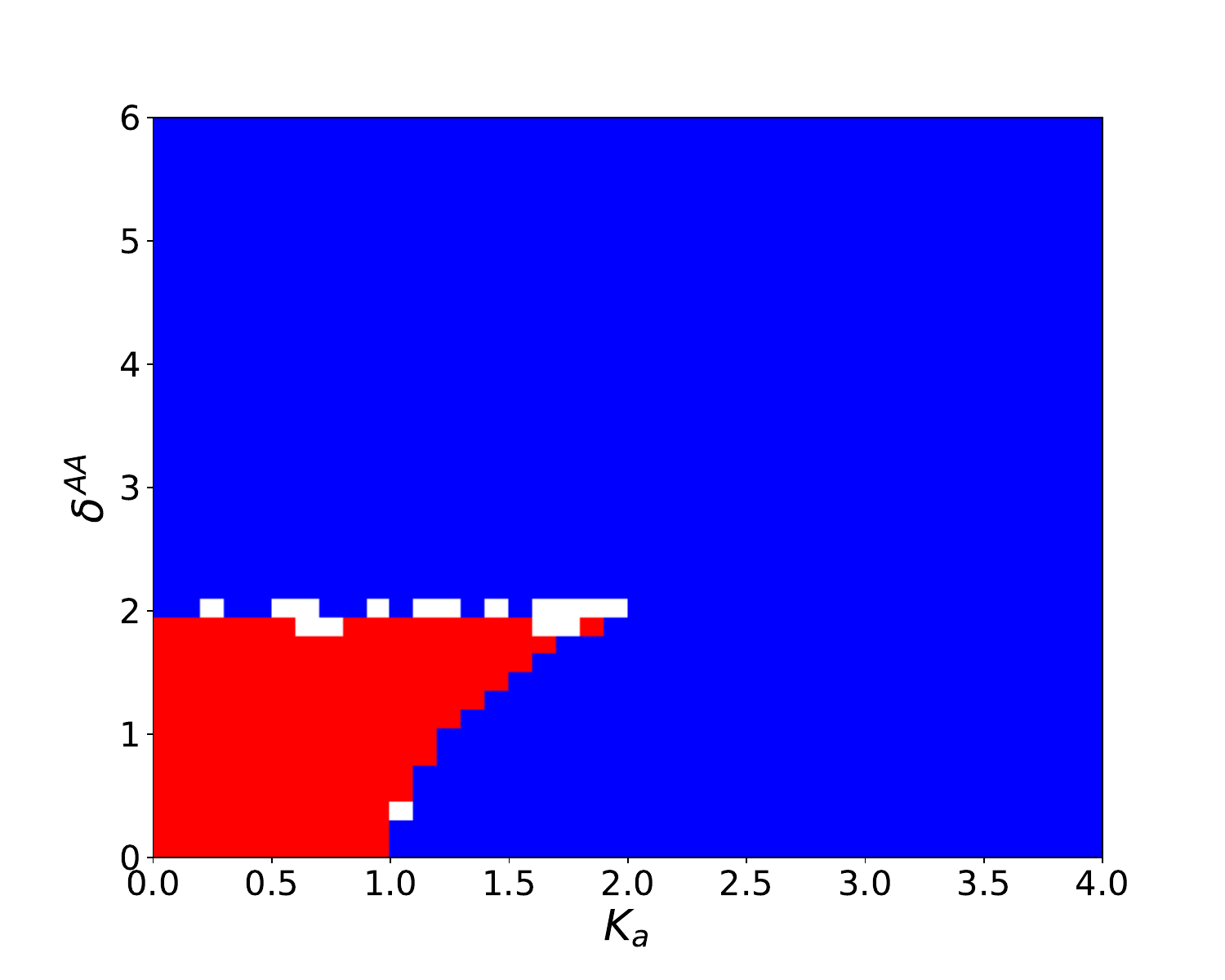}
    \caption{Plot of topological invariant defined in Eq.~(\ref{eq:LTM1}) in the phase space of $K_{a}$ and $\delta^{AA}$. The phase plots clearly shows that at the critical strength of quasi-periodic strength $\delta^{AA}=2$, the topological phase transition is independent of the value of $K_{a}$.}
    \label{fig:e2}
\end{figure}
\begin{figure}[ht!]
    \centering
    \includegraphics[width=\linewidth]{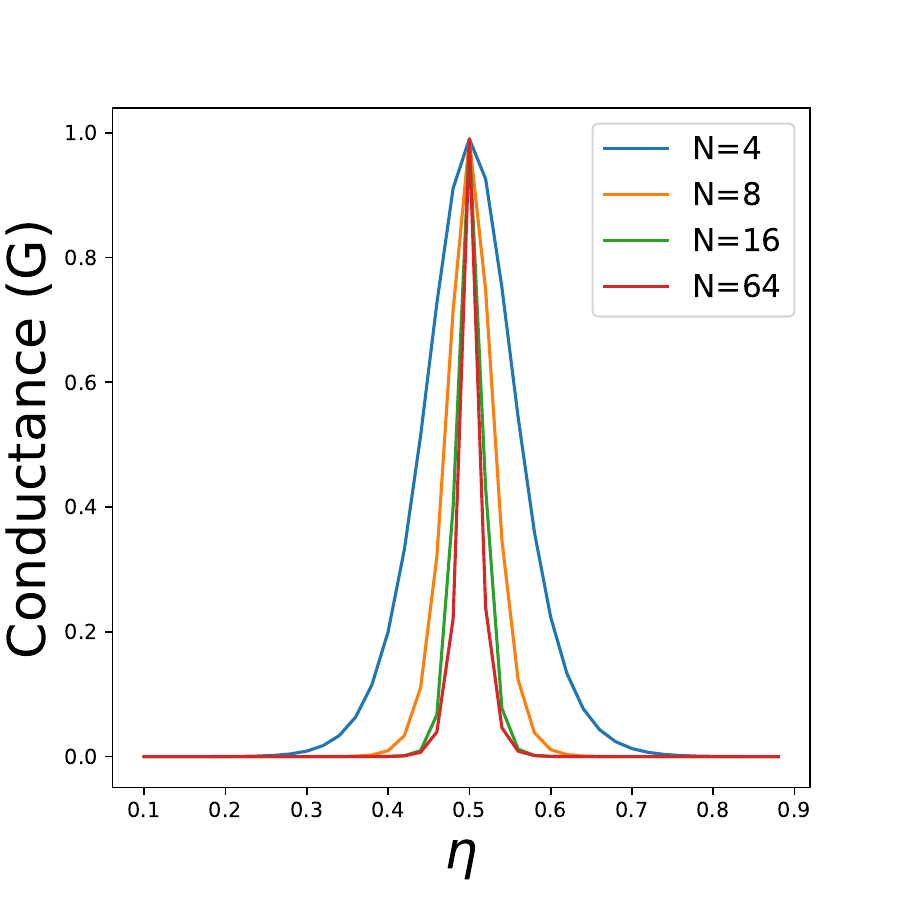}
    \caption{The plot illustrates conductance as a function of the hopping parameter $\eta$, where $t_{1}=\eta$ and $t_{2}=1-\eta$, for the Hamiltonian described in Eq.~\eqref{eq:83}. Conductance is calculated using the Non-Equilibrium Green's Function (NEGF) technique. At $\eta=0.5$, $t_{1}=t_{2}$, representing the metallic state with high conductance, which then fall apart in both sides. The rate of falling reflects the chiral nature of the edge states.}
    \label{fig:e2A}
\end{figure}
\begin{figure}[ht!]
    \centering
    \includegraphics[width=0.7\linewidth]{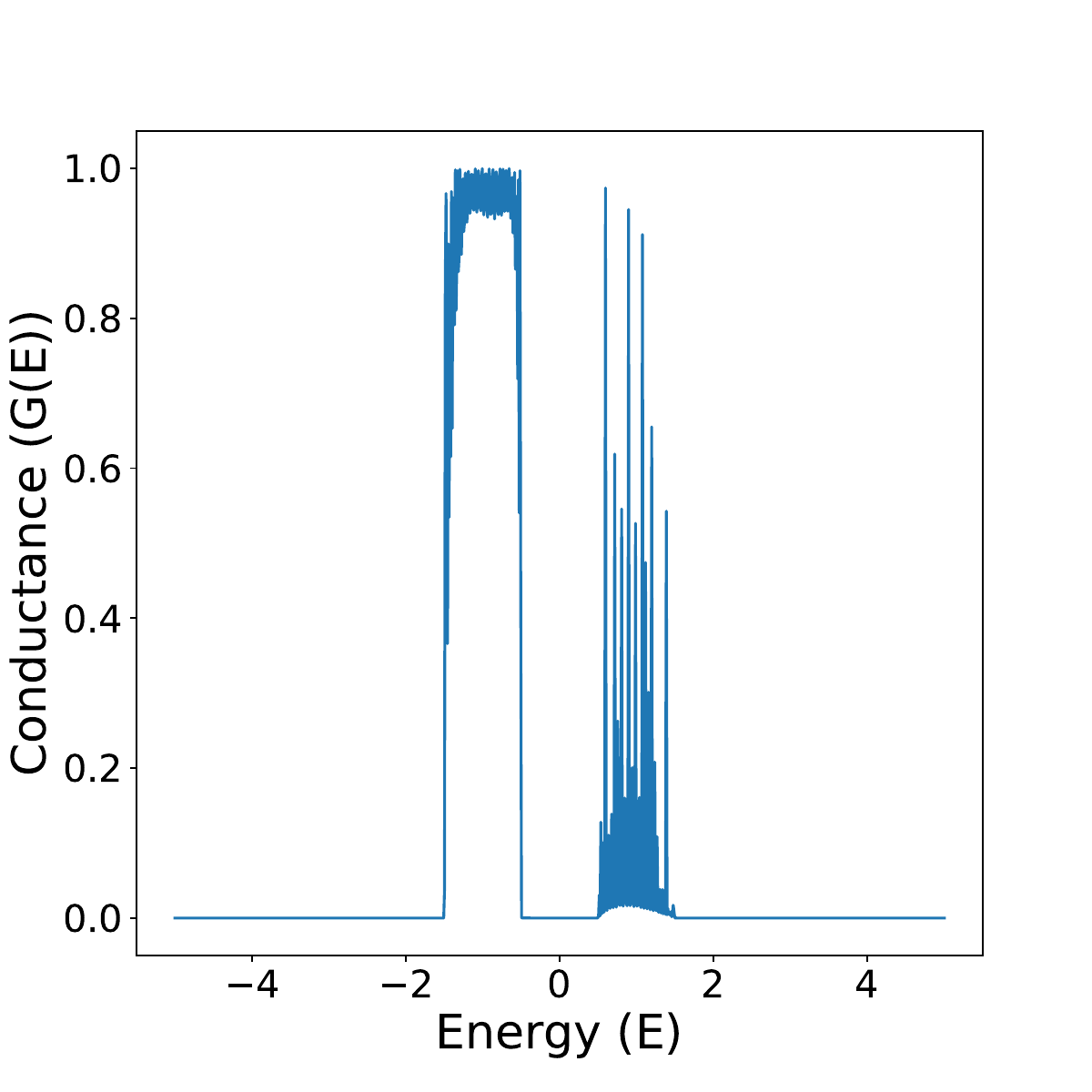}
    \caption{The plot shows conductance in units of ($\frac{e^{2}}{2\pi h}$) versus energy $E$ for the dynamical matrix described in Eq.~\eqref{eq:dynamicalmat1} with open boundaries. Conductance is calculated using the Non-Equilibrium Green's Function (NEGF) technique and becomes quantized to $1$ when the Fermi energy is within the indicated energy range. }
    \label{fig:e2C}
\end{figure}
\begin{figure}[ht!]
    \centering
    \includegraphics[width=\linewidth]{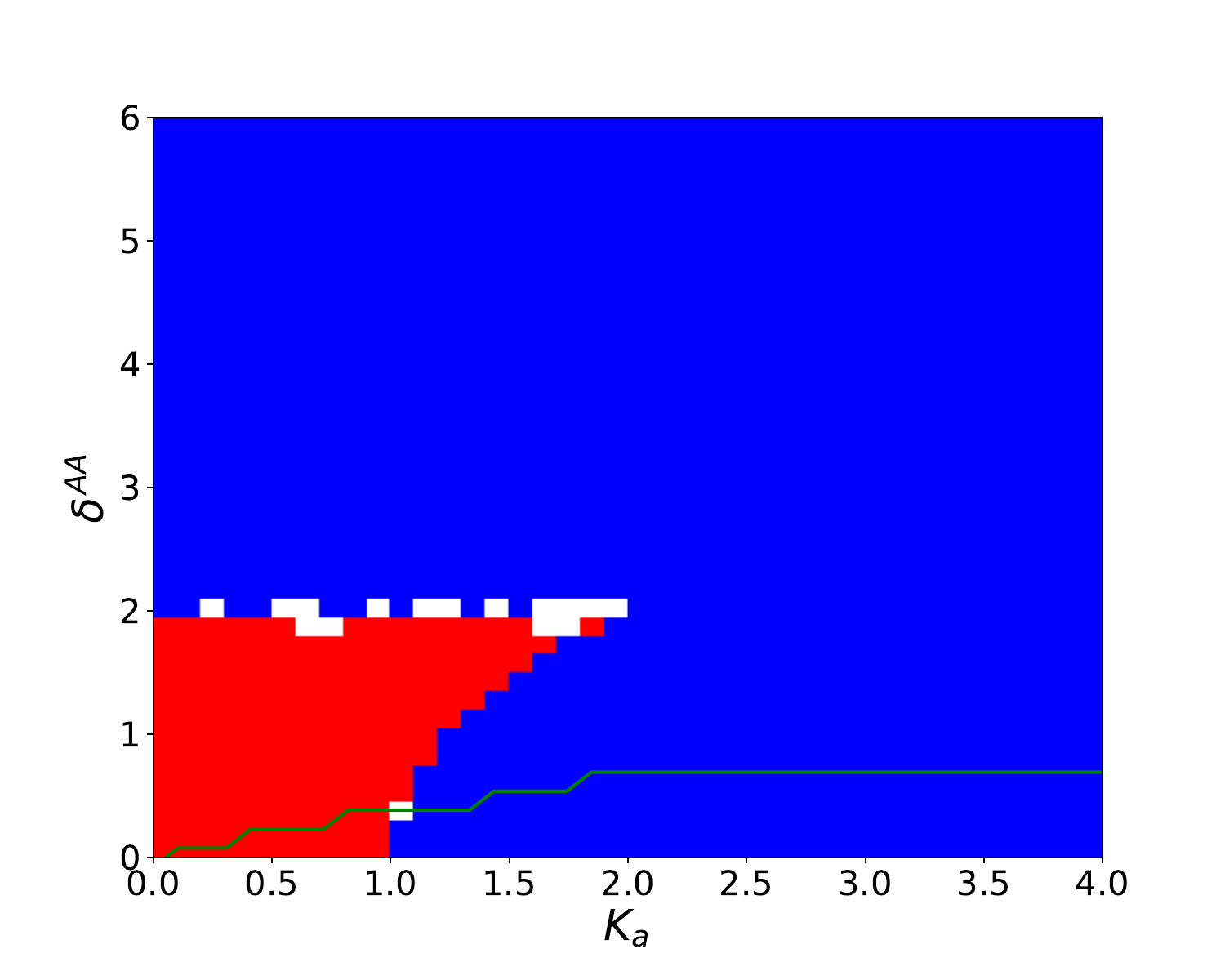}
    \caption{The plot of the topological invariant from Eq.~(\ref{eq:LTM1}) is shown in the phase space of \( K_a \) and \( \delta^{AA} \). The white strip marks the boundary of the phase transition between the non-trivial and trivial topological phases, while the green strip distinguishes the non-localized phase from the localized phase. This figure clearly demonstrates that these two types of phase transitions are uncorrelated.}
    \label{fig:e2B}
\end{figure}
One can identify two masses (having same mass) in each unit cell by a sub-lattice labelling $\alpha={A,B}$ along with a unit cell labelling $j \in [1,N]$. There are two types of spring stiffness constants: $K_{a}$ for intra-cellular and $K_{e}$ for inter-cellular interactions, analogous to the hopping parameters in the original SSH chain. By tuning those parameters in this mechanical setting, different topological phases, akin to the original quantum SSH model, can be realized. The equation of motion for each mass can be formulated using Newton's second law or, equivalently, the Euler-Lagrange equations, which are commonly employed in normal modes analysis. The equations can be compactly expressed in matrix notation, assuming each mass undergoes harmonic displacement over time,
\begin{equation}
    u_{j}(t)=u_{j}e^{-i\omega_{j}t}.
\end{equation}
The second-order equations of motion can be converted into a linear matrix equation. 
\begin{figure}[ht!]
    \centering
    \includegraphics[width=\linewidth]{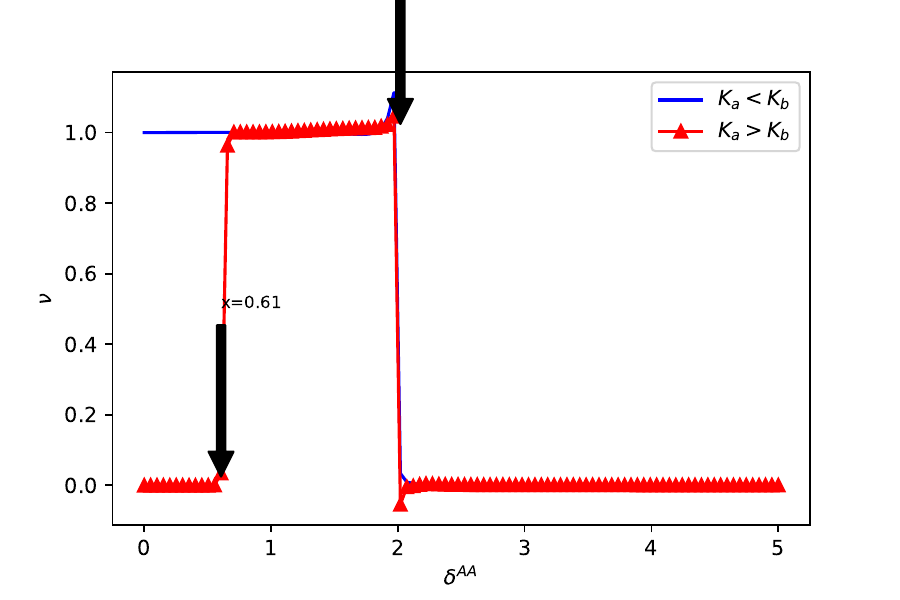}
    \caption{Evolution of topological invariant defined in Eq.~(\ref{eq:LTM1}) versus the quasi-periodic modulation strength $\delta^{AA}$ . The plot shows emergence of TAI phase in some intermediate value of modulation strength $\delta^{AA}$, the exact value is calculated in Eq.~(\ref{eq:Bound1}) and Eq.~(\ref{eq:Bound2}). The black arrows does not indicate the range of TAI phase, it only indicates the values where the LTM does not converge to winding number due to finite size effects.}
    \label{fig:e3}
\end{figure}
This matrix, known as the dynamical matrix, contains the eigenvectors and eigenvalues that represent the normal modes and their frequencies,
\begin{equation}
    DU=\omega^{2}U,
\end{equation}
where $D$ is the dynamical matrix of the system having dimensions $2N\times2N$, and $U$ is displacement vector associated with the displacement of each mass $U=[u_{1}^{A},u_{1}^{B},u_{2}^{A},u_{2}^{B},\cdot \cdot \cdot,u_{N}^{A},u_{N}^{B}]^{T}$ with $u_{j}^{\alpha}$ is the displacement of mass with unit cell index $j$ and sub-lattice index $\alpha=A/B$. The explicit form for the matrix is 
\begin{equation}\label{eq:dynamicalmat1}
  D= \begin{bmatrix}
K_{a}+K_{b} & -K_{a} & \dots & 0 \\
-K_{a} & K_{a}+K_{b} & -K_{b} & \dots \\
\vdots & \vdots & \ddots & -K_{a} \\
\dots & \dots & -K_{a} & K_{a}+K_{b} \\
\end{bmatrix} .
\end{equation}
\begin{figure}[ht!]
    \centering
    \includegraphics[width=\linewidth]{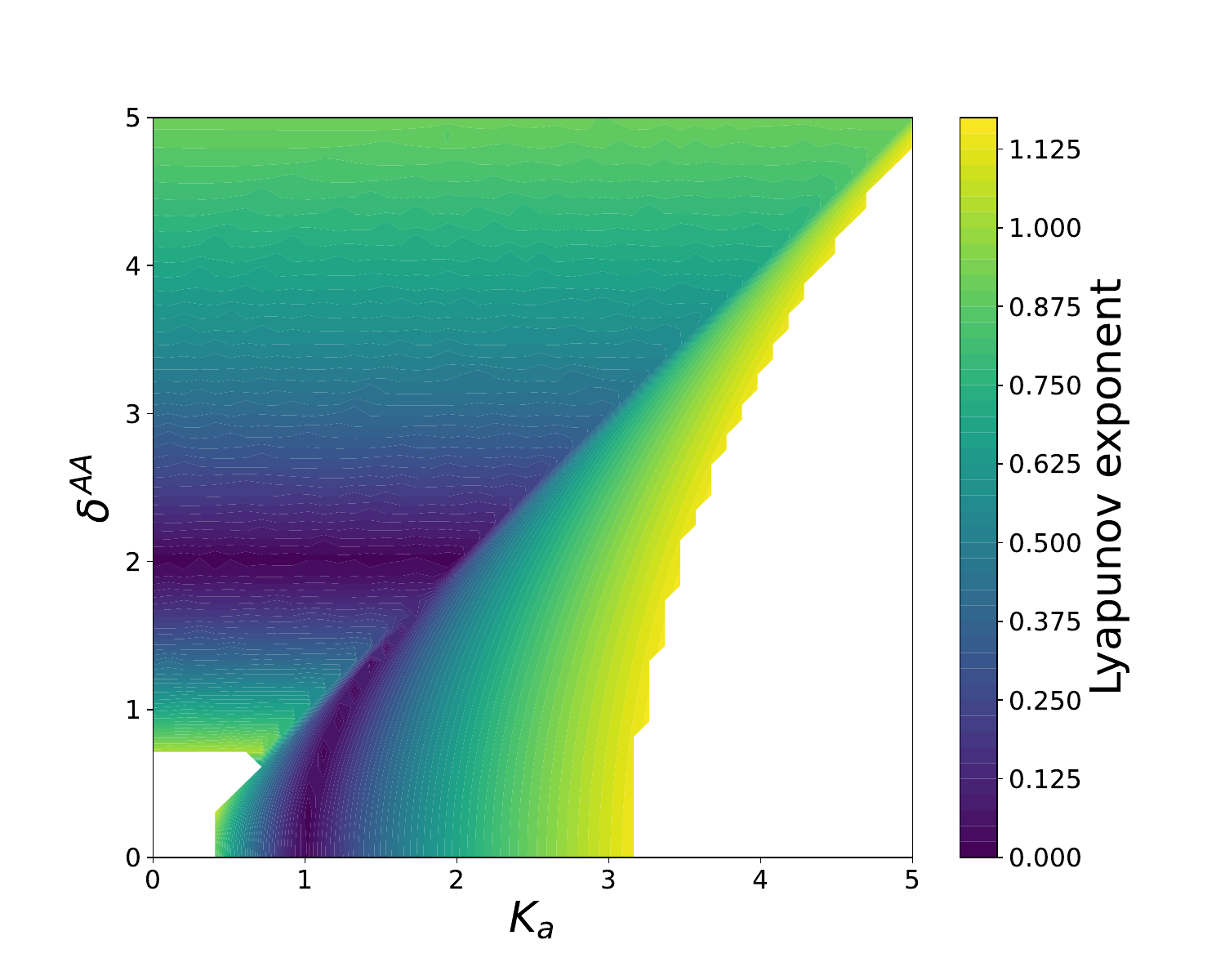}
    \caption{The plot of the Lyapunov exponent in the phase space of $K_{a}$ and $\delta^{AA}$ indicates that a divergence in localization length is marked by the Lyapunov exponent approaching $0$. This boundary of localization length divergence aligns precisely with our phase plot of the topological invariant, as illustrated in Fig.~\ref{fig:e2}.}
    \label{fig:e7}
\end{figure}
The matrix has a tri-diagonal structure. 
The eigenvalues of this matrix for odd dimensions ($N=2k+1$) are:
\begin{equation}\label{eq:81}
    \lambda_{m}=(K_{a}+K_{b})-(+)\sqrt{K_{a}^{2}+K_{b}^{2}+2K_{a}K_{b}\cos(\frac{m\pi}{k+1})},
\end{equation} where $m\in[1,k]$.
For even dimensions ($N=2k$), the eigenvalues are given as: 
\begin{equation}\label{eq:82}
    \lambda_{m}=(K_{a}+K_{b})-(+)\sqrt{K_{a}^{2}+K_{b}^{2}+K_{a}K_{b}z_{m}},
\end{equation} where 
$z_{m}$ can be obtained using the recursion relation involving Chebyshev's polynomial of the first kind, $G_{k}(w)$, as follows:
\begin{equation}
    G_{k}(w)=(\frac{K_{b}}{K_{a}})^{\frac{1}{2}}G_{k-1}(w),
\end{equation} for $w=z_{m}$.
The eigenvalue expression closely resembles the dispersion relation of the original tight-binding SSH Hamiltonian, characterized by the hopping parameters $t_{1}$ (intracellular) and $t_{2}$ (intercellular):
\begin{equation}\label{eq:84}
    E(k)=\sqrt{t_{1}^{2}+t_{2}^{2}+2t_{1}t_{2}\cos(k)},
\end{equation} where $k$ represents a wavevector in the parameter space of the model, which corresponds to $1$st Brillouin zone.
The tight binding hopping Hamiltonian for SSH model is given as :
\begin{widetext}
\begin{equation}\label{eq:83}
    H=\sum_{n}t_{1}(|n,A\rangle \langle n,B|+h.c)+\sum_{n}t_{2}(|n,A\rangle \langle n+1,B|+|n,B\rangle \langle n+1,A|).
\end{equation}
\end{widetext}
\begin{figure*}[htbp]
    \centering
    \includegraphics[keepaspectratio, width=0.45\textwidth]{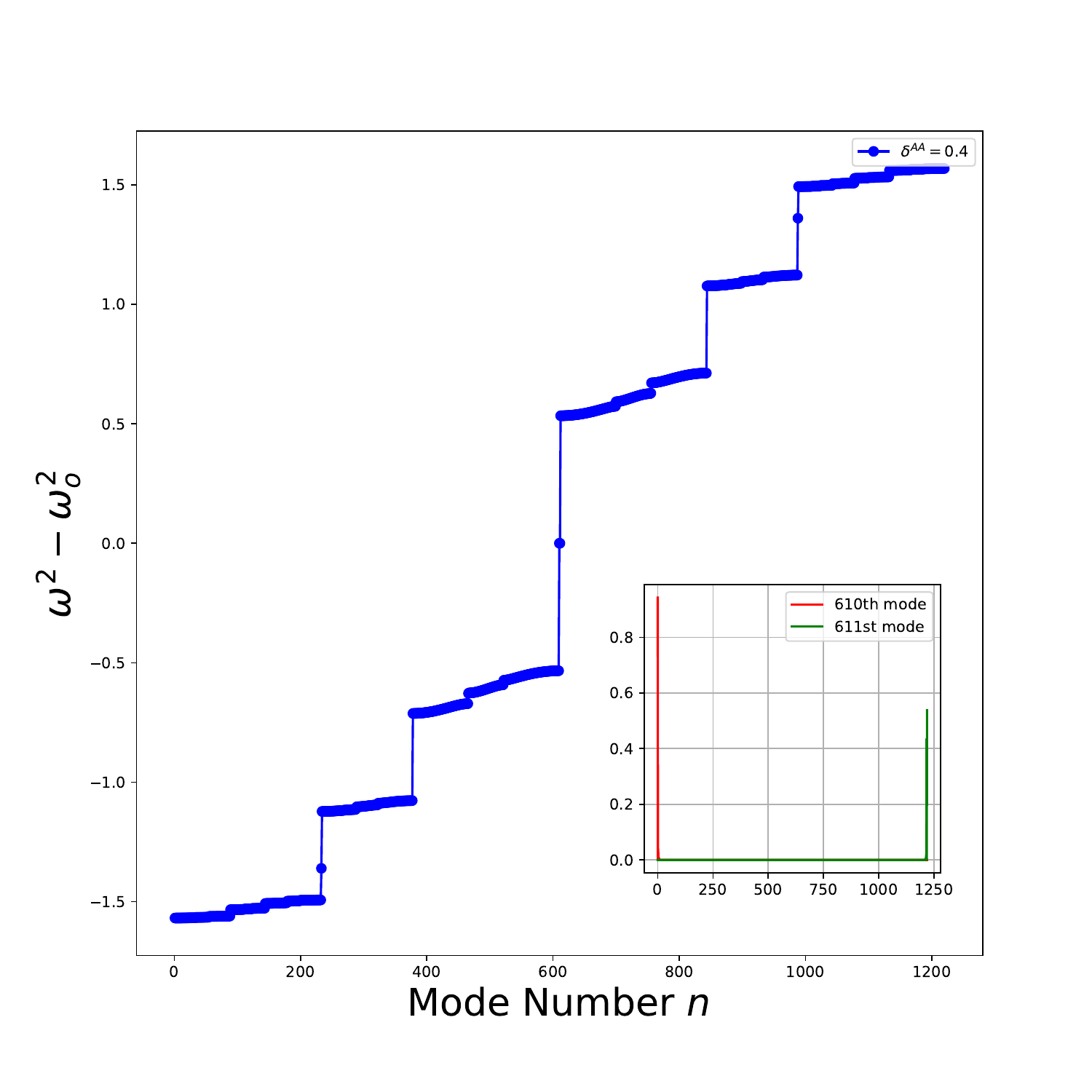}
    \includegraphics[keepaspectratio, width=0.45\textwidth]{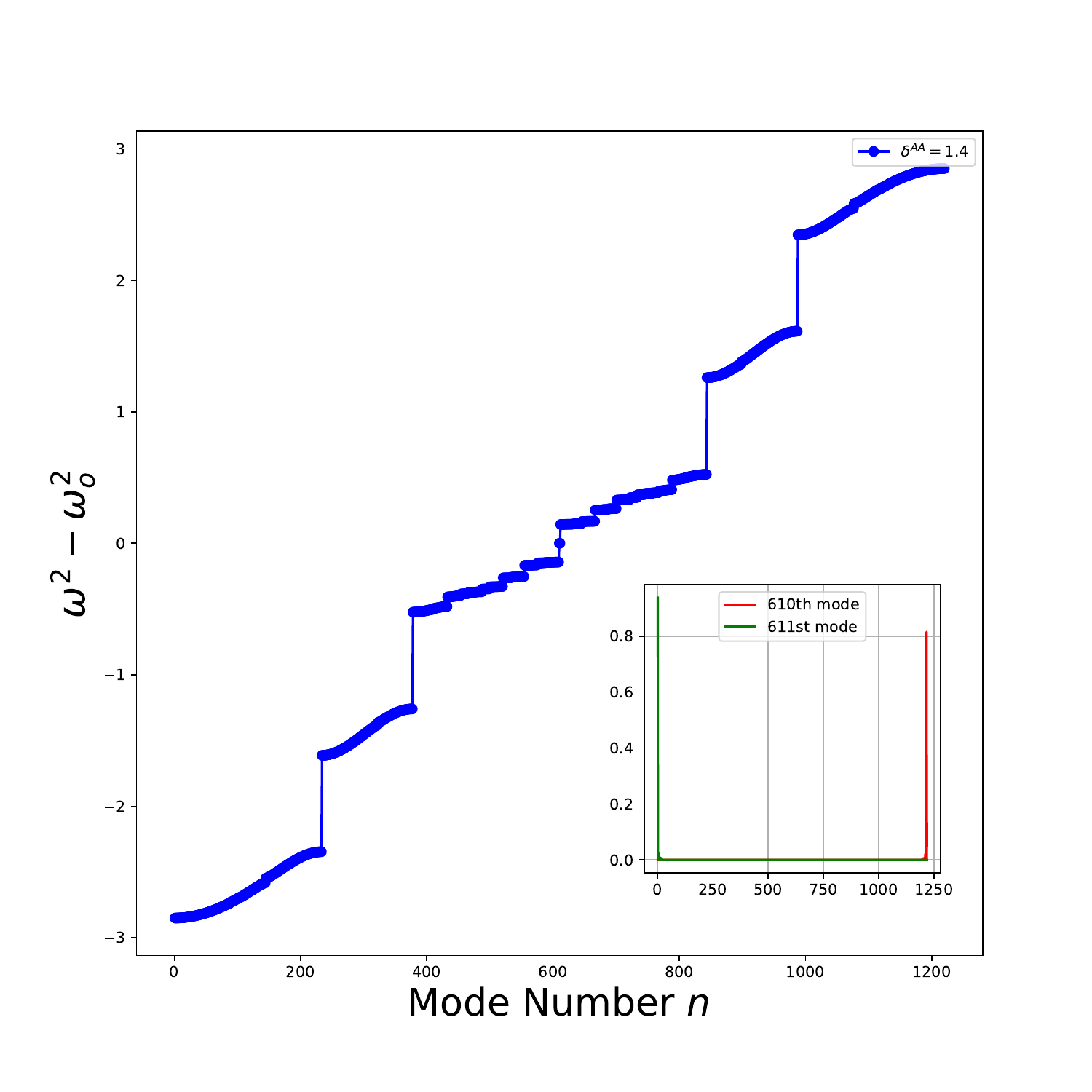}
     \put(-402,160){\textbf{(a) }}
     \put(-170,157){\textbf{(b)}}
     \\
    \includegraphics[keepaspectratio, width=0.45\textwidth]{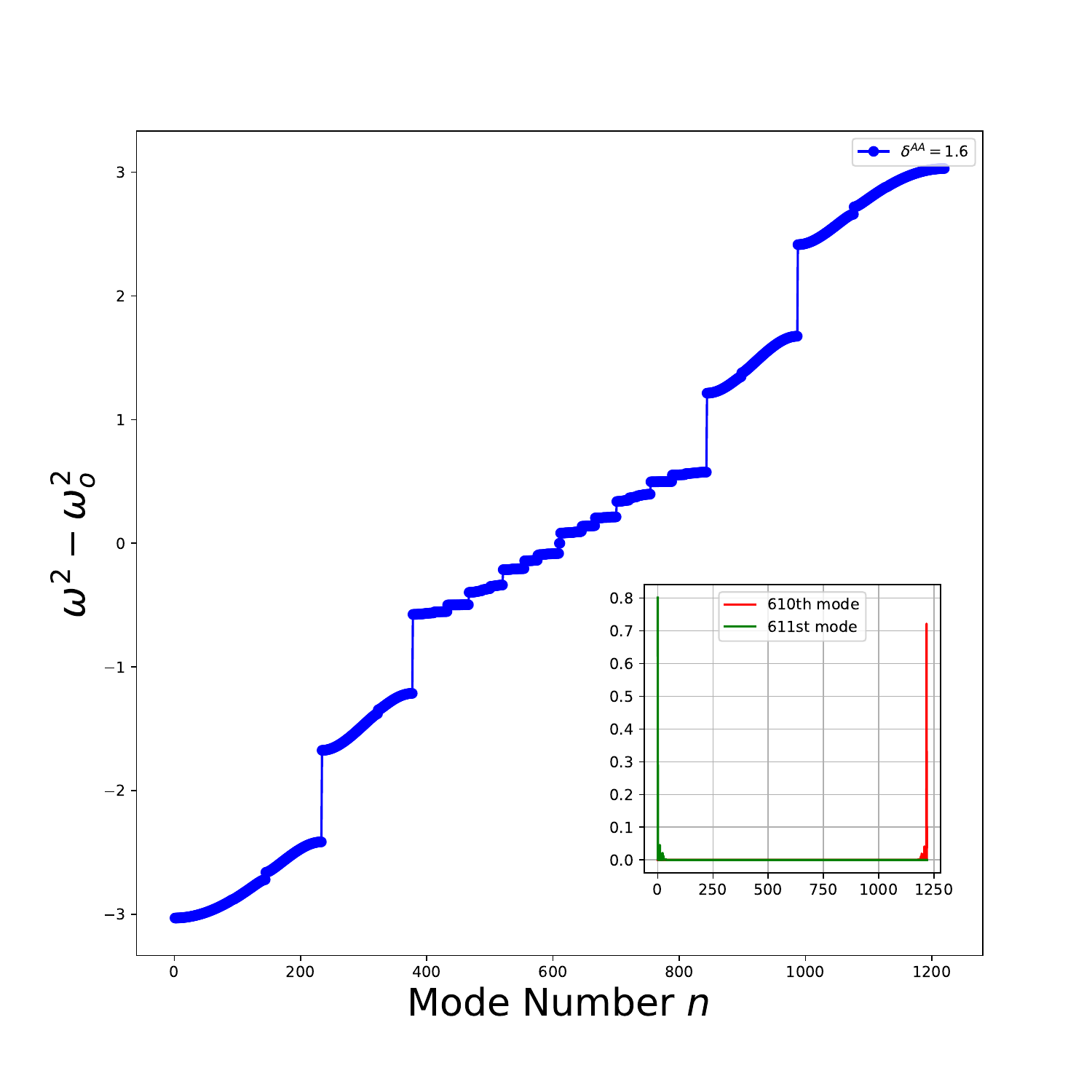}
    \includegraphics[keepaspectratio, width=0.45\textwidth]{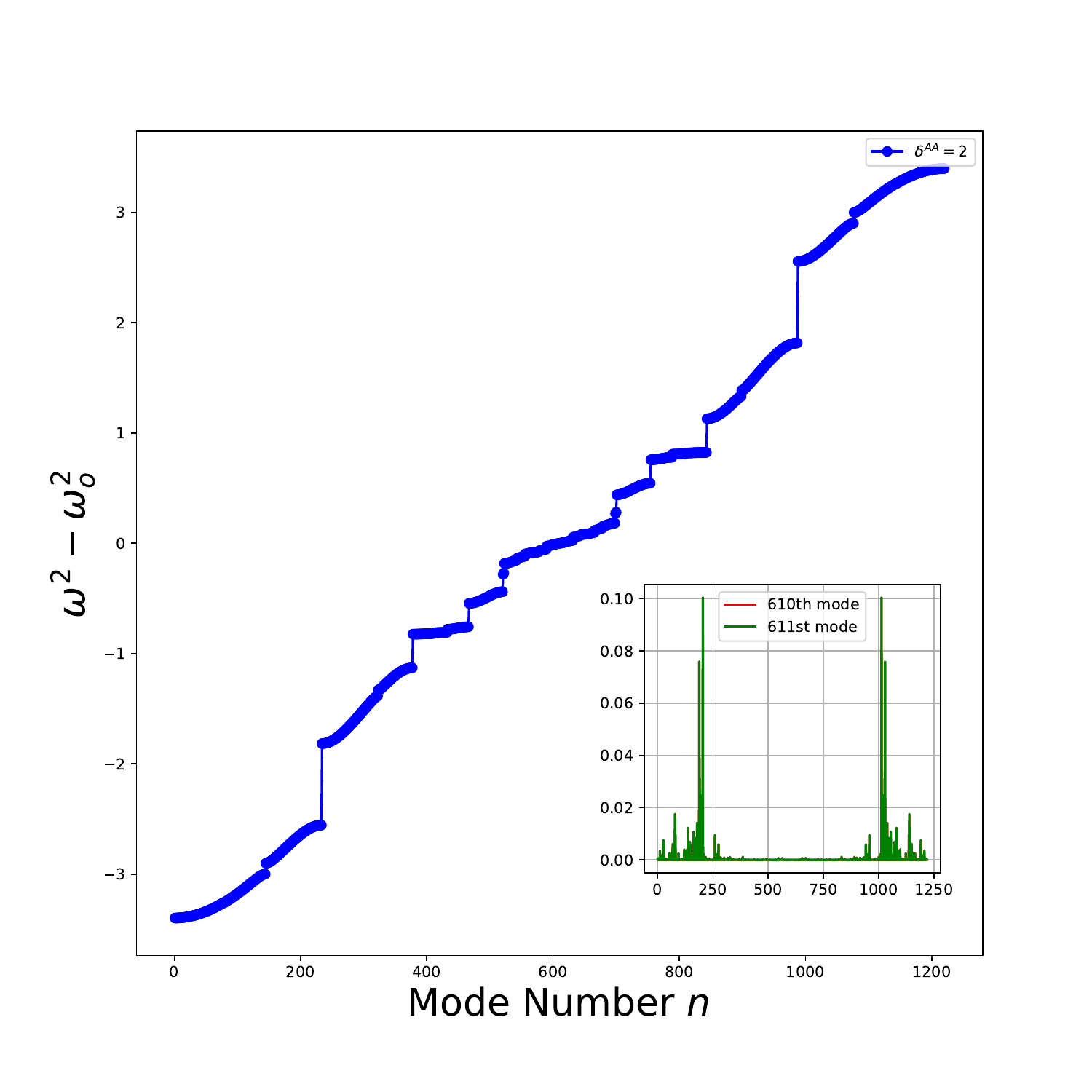}
    \put(-400,160){\textbf{(c) }}
    \put(-170,157){\textbf{(d) }}
    \\
    \caption{Plot of energy versus mode number for a 1D spring mass chain under following conditions: (a) $K_{a}<K_{b}$, (b) $K_{a}>K_{b}$ for $\delta^{AA}=1.4$, (c) $K_{a}>K_{b}$ for $\delta^{AA}=1.6$, (d) $K_{a}>K_{b}$ for $\delta^{AA}=2$. For all cases we calculate the eigenvalues using the dynamical matrix from Eq.~(\ref{eq:dynamicalmat1}). For cases (a), (b), (c) the spectrum  shows a gap near the Fermi level with mid-gap modes. For those cases, the insets displays the probability distribution of the mid-gap states, which are localized at the edges. The color will indicate the chirality of two edge states, highlighting the topologically non-trivial phase. In case (d), the plot  reveals an absence of mid-gap modes in the spectrum. The inset  illustrates the probability distribution of mid-gap states extending across the bulk without any presence of edge modes. This phenomenon  accompanies lack of chirality, which indicates that our system is indeed in a  phase, which has a non-zero winding number. The scenarios in cases (b) and (c) falls under the TAI phase range, which is shown in Fig.~\ref{fig:e2},  that reveals the presence of conducting edge states protected by the symmetries of the dynamical matrix. In case (d), we verify that as $\delta^{AA}=2$, the TAI phase disappears, leading to a trivial Anderson insulator phase.}
    \label{fig:combined_vector_plot1}
\end{figure*}
Using the equation Eq.~\eqref{eq:83}, by diagonalizing the Hamiltonian in Fourier space and taking advantage of translational invariance, we can easily show that the dispersion relation is given by equation Eq.~\eqref{eq:84}. Conductance for this model is computed using the non-equilibrium Green's function technique, as illustrated in Fig.~\ref{fig:e2A}.
The expression for eigenvalues in Eq.~\eqref{eq:81} and Eq.~\eqref{eq:82} provides motivation for accessing the topological phases in this mechanical model using a real space topological invariant.The details of the calculation and the form for eigenvectors can be found in Appendix.~\ref{sec:appendixB}.
Unlike the initial SSH Hamiltonian, changes in inter-cellular hopping only affect the off-diagonal elements, while in our mechanical scenario, adjustments to the inter-cellular stiffness constant also impact the diagonal elements. However, since all diagonal elements are identical, the system maintains chiral symmetry; these diagonal elements merely shift the eigenvalue spectrum to $\omega_{0}^{2}=K_{a}+K_{b}$. We measure all other eigenvalues relative to this value. We can define the matrix $D^{'} = D - \omega_{0}^{2}I$, that anti-commutes with the chiral operator $\Gamma$, centering the eigenvalue spectrum around $\omega_{0}^{2}=0$, as shown below:
\begin{equation}
    \Gamma(D-\omega_{0}^{2}I)+(D-\omega_{0}^{2}I)\Gamma=0,
\end{equation}
where $I$ is the identity matrix having dimension $2N\times2N$, and $\Gamma$ is the chiral operator, whose matrix representation is:
{\small
\begin{equation}\label{eq:chiralmatrix}
  \Gamma=  \begin{bmatrix}
1 & 0 & \dots & 0 \\
0 & -1 & 0 & \dots \\
\vdots & \vdots & \ddots & 0 \\
\dots & \dots & 0 & 1 \\
\end{bmatrix} .  
\end{equation}
}

\subsection{Characterizing the topology in real space}
In quantum tight-binding hopping models, we compute the winding number in momentum space (where translational invariance is preserved) for topological characterization \cite{CHEN201822}. In large systems, translational symmetry holds in the bulk but is broken near the boundary. In the clean limit (without disorder), our system maintains this symmetry in the bulk, but disorder disrupts it. We will conduct the analysis in real space \cite{PhysRevResearch.3.033012,PhysRevLett.113.046802} and characterize the topology using local topological marker (LTM) \cite{Meier_2018,sircar2024disorderdriventopologicalphase}. An LTM is defined for each unit cell, which converges to the winding number computed using translational symmetry in the periodic system, when averaged away from the boundaries. The primary advantage of the LTM is its ability to handle disordered systems \cite{PhysRevLett.113.046802,PhysRevResearch.3.033012}. For computing LTMs for a chain composed of $N$ unit cells, we need to create two $N\times N$ matrices of eigenvectors from the dynamical matrix: $U_{-} = [U_{1}, U_{2}, \dots, U_{N}]$ for eigenvectors below the Fermi level, and $U_{+} = [U_{N+1}, U_{N+2}, \dots, U_{2N}]$ for those above it. Using these matrices, we can build the projectors for the bands as $P_{-} = U_{-} U_{-}^{T}$ and $P_{+} = U_{+} U_{+}^{T}$, allowing us to construct a flat-band Hamiltonian \cite{PhysRevResearch.3.033012}. The flat-band Hamiltonian, topologically equivalent to $D$, is expressed as $Q = P_{+} - P_{-}$. This can be spectral decomposed into sub-lattices $A$ and $B$ as $Q = Q_{AB} + Q_{BA} = \Gamma_{A}Q\Gamma_{B} + \Gamma_{B}Q\Gamma_{A}$, with 
\begin{figure}[ht!]
    \centering
    \includegraphics[width=\linewidth]{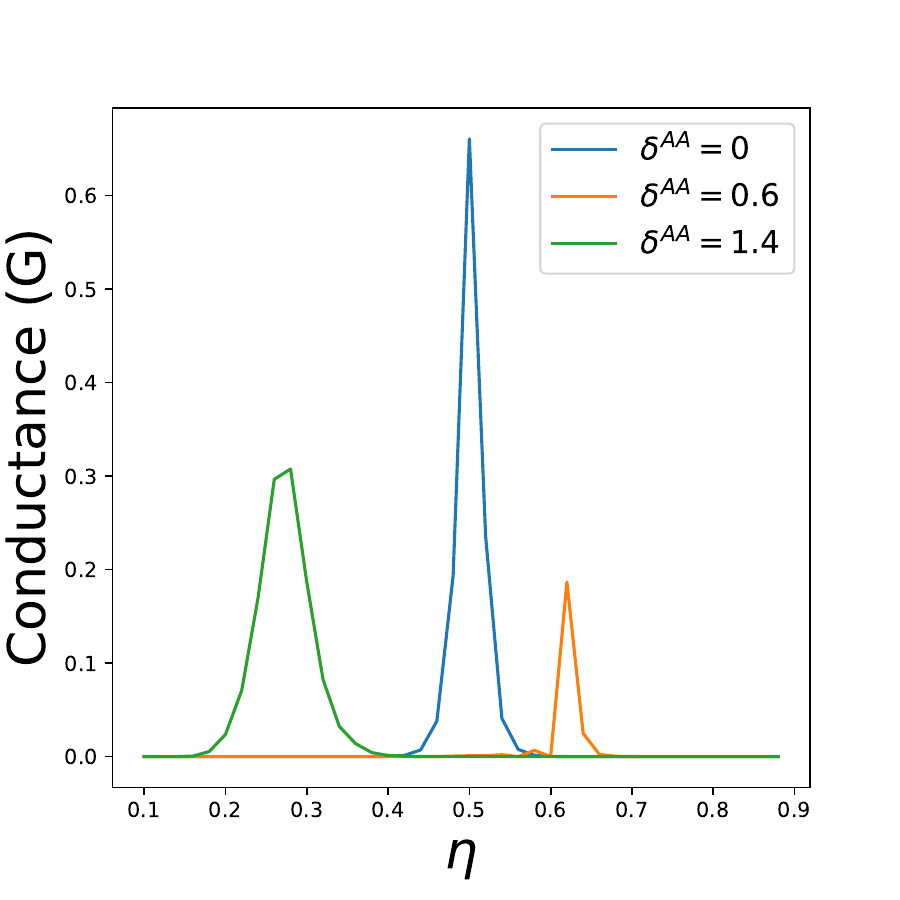}
    \caption{The plot illustrates conductance as a function of the hopping parameter $\eta$, where $K_{a}=\eta$ and $K_{b}=1-\eta$, for the dynamical matrix described in Eq.~\eqref{eq:dynamicalmat2}. Conductance is calculated using the Non-Equilibrium Green's Function (NEGF) technique. At $\eta=0.5$, $K_{a}$ equals $K_{b}$, for $\delta^{AA}=0$, representing the metallic state with high conductance. The calculation uses the rational approximation $Q=\frac{21}{34}$ for a system size of $N=68$.}
    \label{fig:e2B}
\end{figure}
\begin{equation}
    \Gamma_{A/B}= \begin{bmatrix}
1/0 & 0 & 0 & 0 & \dots\\
0 & 0/1 & 0 & 0 & \dots \\
0 & 0 & 1/0 & 0 & \dots\\
\dots & \dots & \dots & \dots & \dots \\ 
\end{bmatrix}   ,
\end{equation}
are the spectral decomposition operators and $\Gamma=\Gamma_{A}-\Gamma_{B}$ is the chiral operator. The mathematical definition of LTM \cite{Meier_2018,PhysRevResearch.3.033012,PhysRevB.109.195427,sircar2024disorderdriventopologicalphase} is
\begin{equation}\label{eq:LTM1}
    \nu(k)=\frac{1}{2}\sum_{\alpha=A,B}{(Q_{BA}[X,Q_{AB}])_{k\alpha,k\alpha}+(Q_{AB}[Q_{BA},X])_{k\alpha,k\alpha}} ,
\end{equation}
$X$ is defined as the position operator, whose matrix representation has dimensions $ 2N \times 2N$. The specific form of $X$ is given by $X = [-N, -N, -(N-1), -(N-1), \dots, (N-1), (N-1)]$, all arranged in as diagonal elements with off-diagonal components being set to $0$. The local topological marker (LTM) converges to the winding number and can have only two values: $\nu = 0$, characterizes topologically trivial phase, while $\nu=1$ characterizes topologically non-trivial phase, respectively. Each phase is tuned to a specific ratio between the parameters $K_{a}$ and $K_{b}$, as shown in Fig.~\ref{fig:combined_vector_plot1}. The ratio $\frac{K_{b}}{K_{a}} > 1$ corresponds to the non-trivial phase, which is characterised by the system possessing conducting edge states, while $\frac{K_{b}}{K_{a}} < 1$ corresponds to the trivial phase, that does not host conducting edge states.

\section{MECHANICAL MODEL MODULATED WITH INTRACELL AUBRY-ANDRE SPRING CONSTANTs}\label{sec:3}
\subsection{Dynamical matrix of the modulated spring-mass system}
Our model is a nearest-neighbour mechanical model \cite{PhysRevResearch.3.033012,PhysRevB.109.195427,sircar2024disorderdriventopologicalphase} that is influenced by $AA$ modulation. The $AA$ modulation is defined by a sinusoidal term whose periodic characterisation is in-commensurate with the underlying lattice structure, because of the irrational number $\beta$, defined in our case as the inverse of golden ratio, $g^{-1}=\frac{2}{1+\sqrt{5}}$. 
\begin{figure*}[htbp]
    \centering
    \includegraphics[keepaspectratio, width=0.45\textwidth]{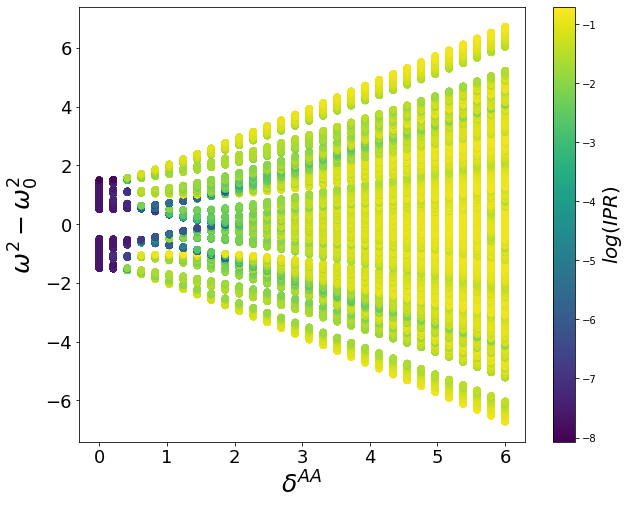}
    \includegraphics[keepaspectratio, width=0.45\textwidth]{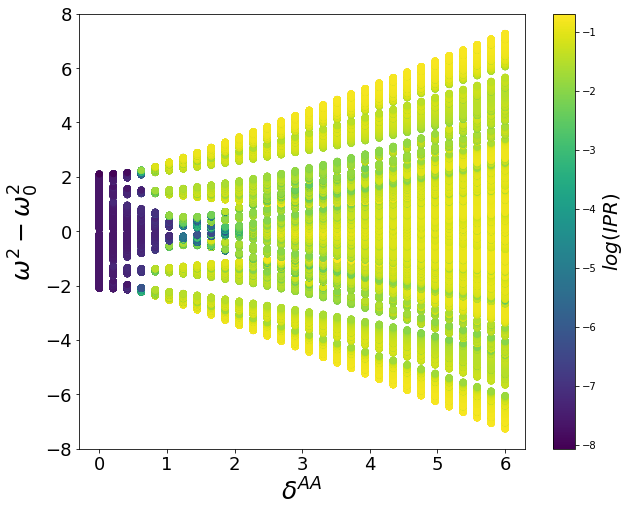}
     \put(-402,160){\textbf{(a) }}
     \put(-170,157){\textbf{(b)}}
     \\
    \includegraphics[keepaspectratio, width=0.45\textwidth]{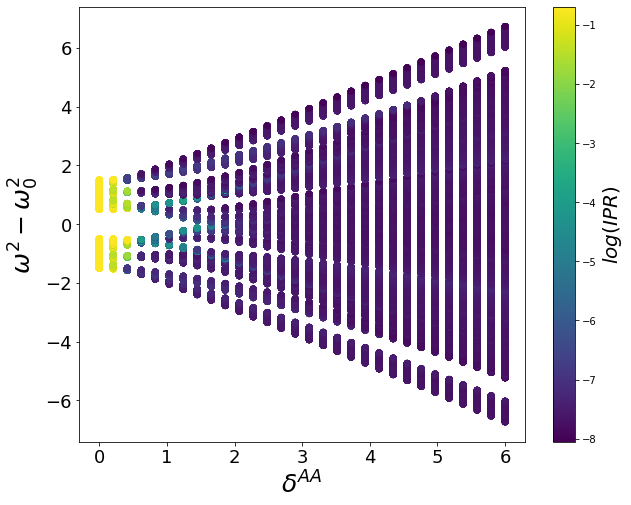}
    \includegraphics[keepaspectratio, width=0.45\textwidth]{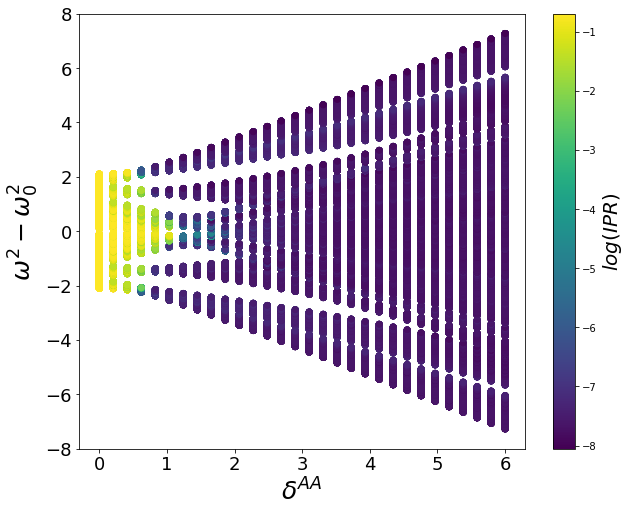}
    \put(-400,160){\textbf{(c) }}
    \put(-170,157){\textbf{(d) }}
    \\
    \caption{Plot of $\ln(IPR)$ as a function of the eigenvalue spectrum $\omega^{2}-\omega_{o}^{2}$ for $N=1597$ unit cells, $\phi=0$. In real space, the transition of eigenstates take place from extended to localized, and from localized to extended in momentum space. For $K_{a}>K_{b}$, one can observe the presence of critical states with in the parameter regime of $\delta^{AA}$ between $1$ and $2$. In this case the eigenstates are either localized or in critical phase for real space, while extended or in critical phase for momentum space.This provides an indication of the presence of a TAI phase, in this parameter regime, which we have quantitatively verified in Fig.~\ref{fig:combined_vector_plot3}, Fig.~\ref{fig:combined_vector_plot4}. (a): $K_{a}<K_{b}$ (real space), (b): $K_{a}>K_{b}$ (real space), (c): $K_{a}<K_{b}$ (momentum space), (d): $K_{a}>K_{b}$ (momentum space).}
    \label{fig:combined_vector_plot2}
\end{figure*}
The equations of motion of our model for type $A$ mass in $j$th unit cell will be 
\begin{equation}
    \begin{split}
   \ddot{u}^{A}_{j} &= -[K_{a} +  K_{b} + \delta^{AA} \cos(2\pi \beta j + \phi)] u^{A}_{j} \\
   &\quad + [K_{a}+\delta^{AA} \cos(2\pi \beta j + \phi)] u^{B}_{j} + [K_{b}] u^{B}_{j-1}.
\end{split}
\end{equation}
\begin{figure*}[htbp]
    \centering
    \includegraphics[keepaspectratio, width=0.45\textwidth]{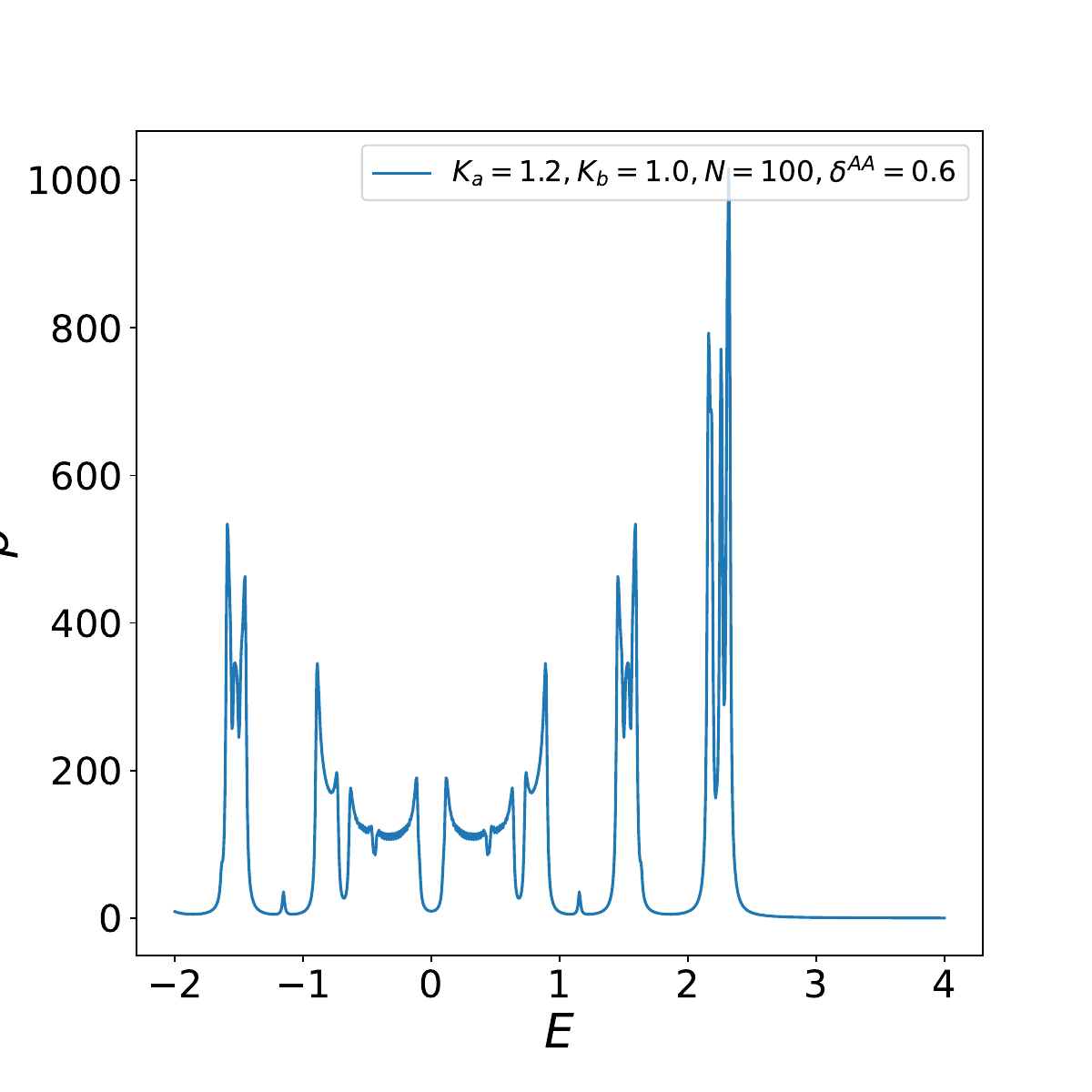}
    \includegraphics[keepaspectratio, width=0.45\textwidth]{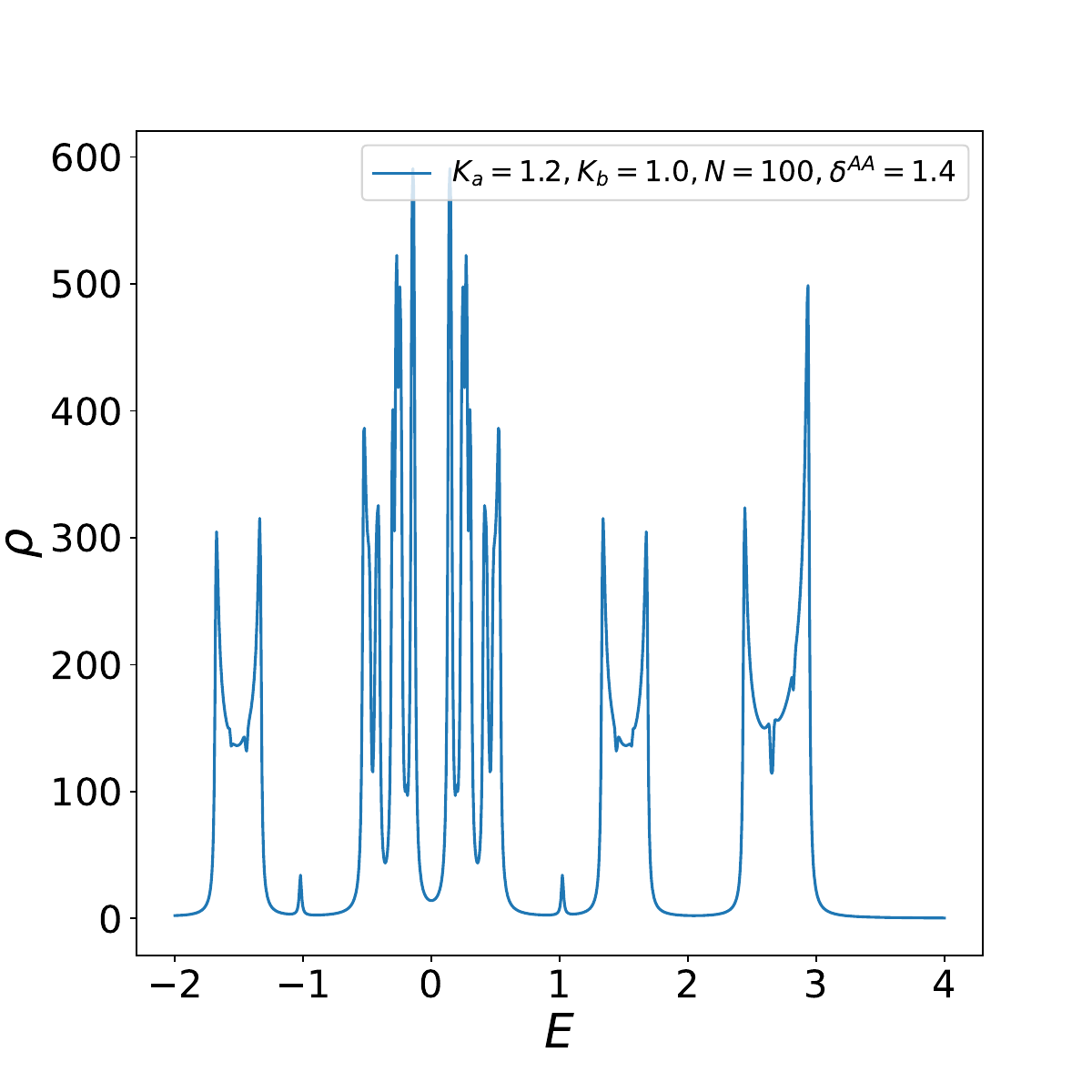}
     \put(-410,160){\textbf{(a) }}
     \put(-190,157){\textbf{(b)}}
     \\
    \includegraphics[keepaspectratio, width=0.45\textwidth]{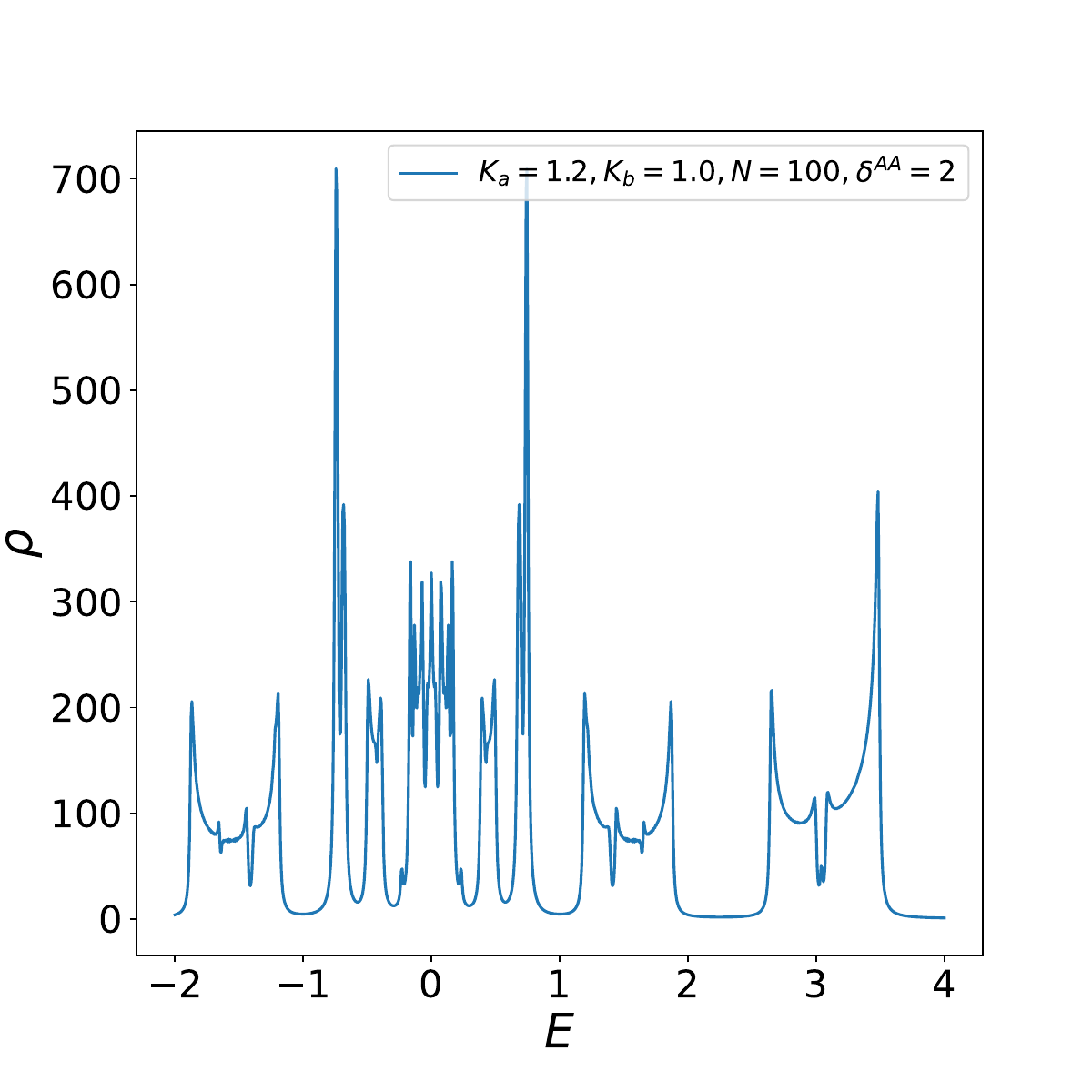}
    \includegraphics[keepaspectratio, width=0.45\textwidth]{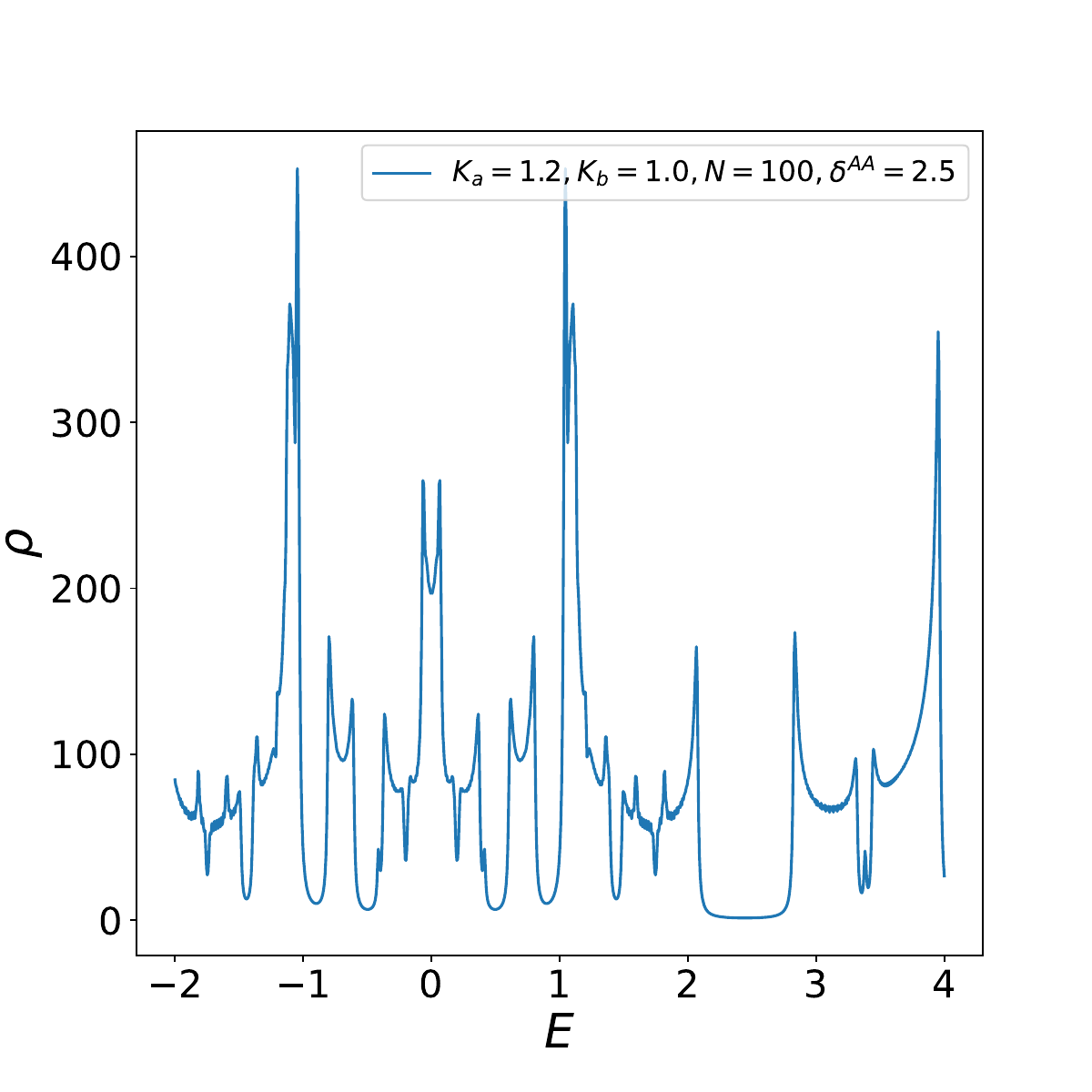}
    \put(-410,160){\textbf{(c) }}
    \put(-190,157){\textbf{(d) }}
    \\
    \caption{Plot of local density of states with respect to energy for the parameter values $K_{a}>K_{b}$ under periodic boundary conditions (PBC). (a): $\delta^{AA}=0.6$, (b): $\delta^{AA}=1.4$, (c): $\delta^{AA}=2.0$, (d):$\delta^{AA}=2.5$. The plot (a), (b) shows a region near the zero energy level, where the local density of states vanishes indicating a gap in the bulk spectrum. The cases (a) and (b) are for TAI phase.}
\end{figure*}
Similarly for type $B$ mass in $j$th unit cell, 
\begin{equation}
    \begin{split}
   \ddot{u}^{B}_{j} &= -[K_{a} + K_{b} + \delta^{AA} \cos(2\pi \beta (j) + \phi)] u^{B}_{j} \\
   &\quad + [K_{a}+\delta^{AA} \cos(2\pi \beta j + \phi)] u^{A}_{j} + [K_{b}] u^{A}_{j+1}.
\end{split}
\end{equation}
where $\delta^{AA}$ is the $AA$ amplitude (with $j\in[1,N]$ as unit cell index), the spring constants are defined as 
\begin{equation}
    k_{j}=k_{j}+\delta_{j}^{AA}=K_{a}+\delta^{AA}\cos(2\pi \beta j+\phi),
\end{equation} for $j$ even.

Different values of $\phi$ are possible; however, in this article, we will fix $\phi$ at $0$ when discussing incommensurate modulation. One can also use an ensemble of different $\phi$ values and then compute a disorder average of all relevant quantities. In this article, as we have already fixed the value of $\phi$, hence no disorder averaging is needed. Perturbing the stiffness of the springs introduces new diagonal terms, which do not appear in the original SSH model. This perturbation alters both diagonal and off-diagonal terms, hence disrupting the chirality of the dynamical matrix. To maintain the chiral symmetry, we introduce local springs to each mass.
The equations of motion of our model for type $A$ mass in $j$th unit cell will be 
\begin{equation}
    \begin{split}
   \ddot{u}^{A}_{j} &= -[K_{a}  + K_{b} + \delta^{AA} \cos(2\pi \beta j + \phi)+K^{0,A}_{j}] u^{A}_{j} \\
   &\quad + [K_{a} + \delta^{AA} \cos(2\pi \beta j + \phi)] u^{B}_{j} + [K_{b}] u^{B}_{j-1}.
\end{split}
\end{equation}

\begin{figure*}[htbp]
    \centering
    \includegraphics[keepaspectratio, width=0.45\textwidth]{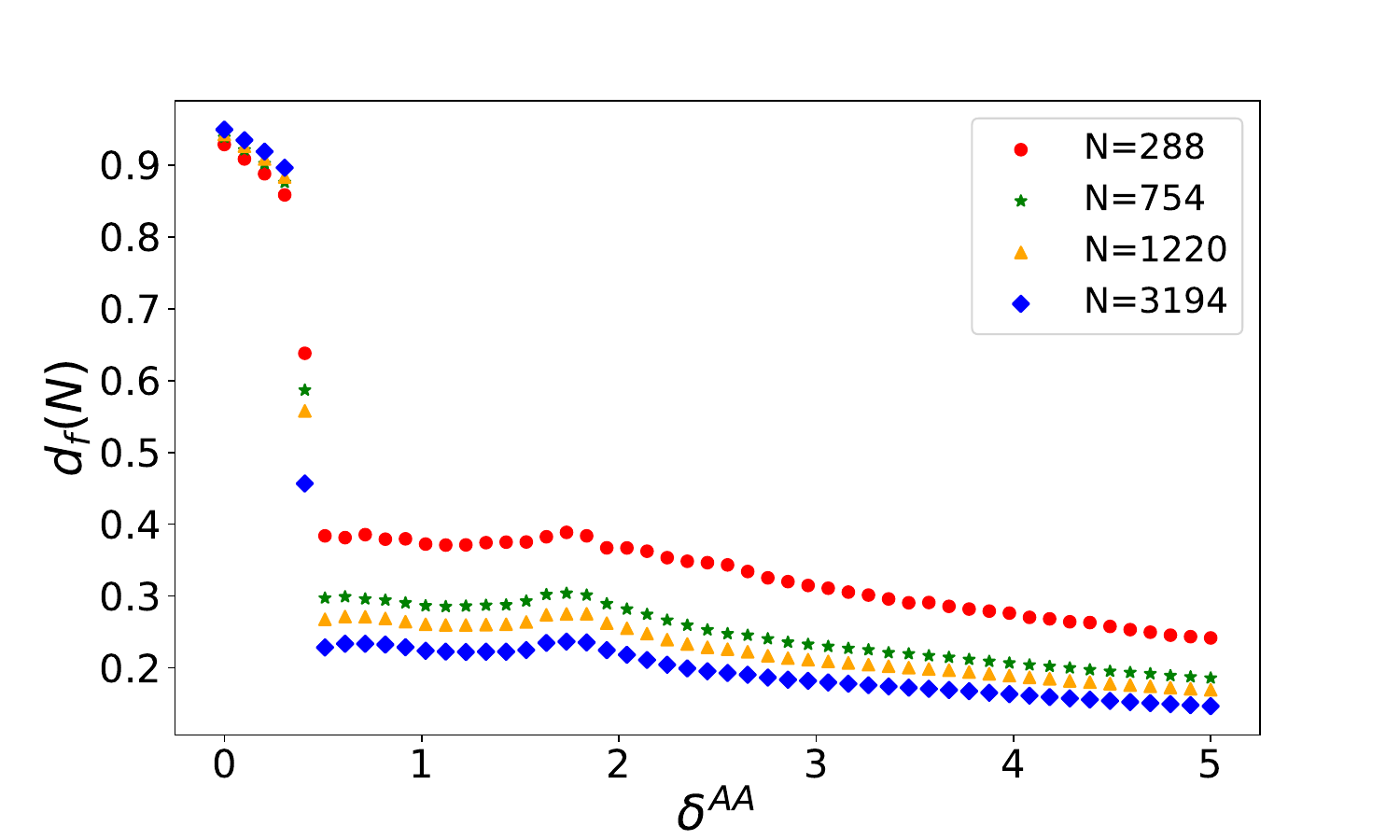}
    \includegraphics[keepaspectratio, width=0.45\textwidth]{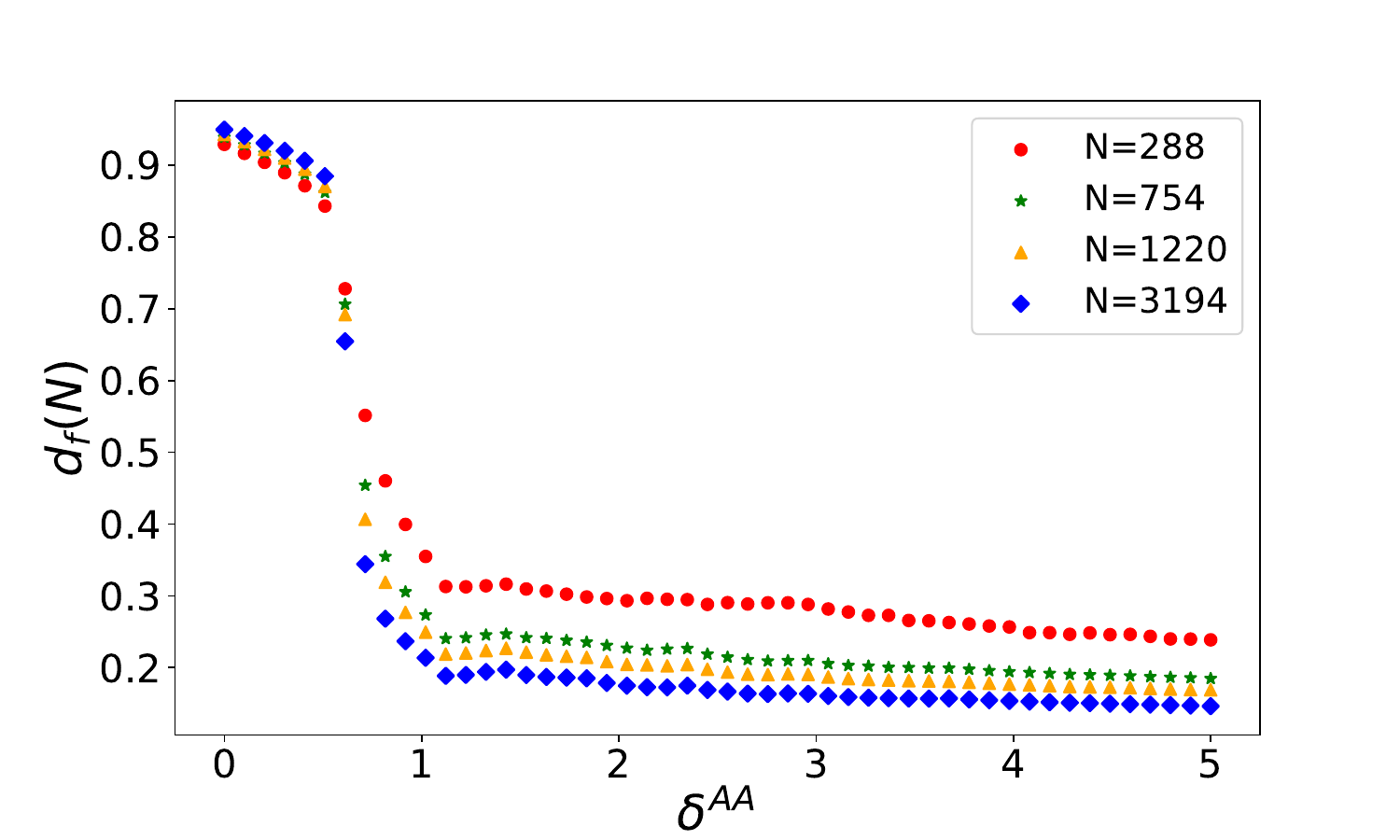}
     \put(-360,110){\textbf{(a) }}
     \put(-145,110){\textbf{(b)}}
     \\
    \includegraphics[keepaspectratio, width=0.45\textwidth]{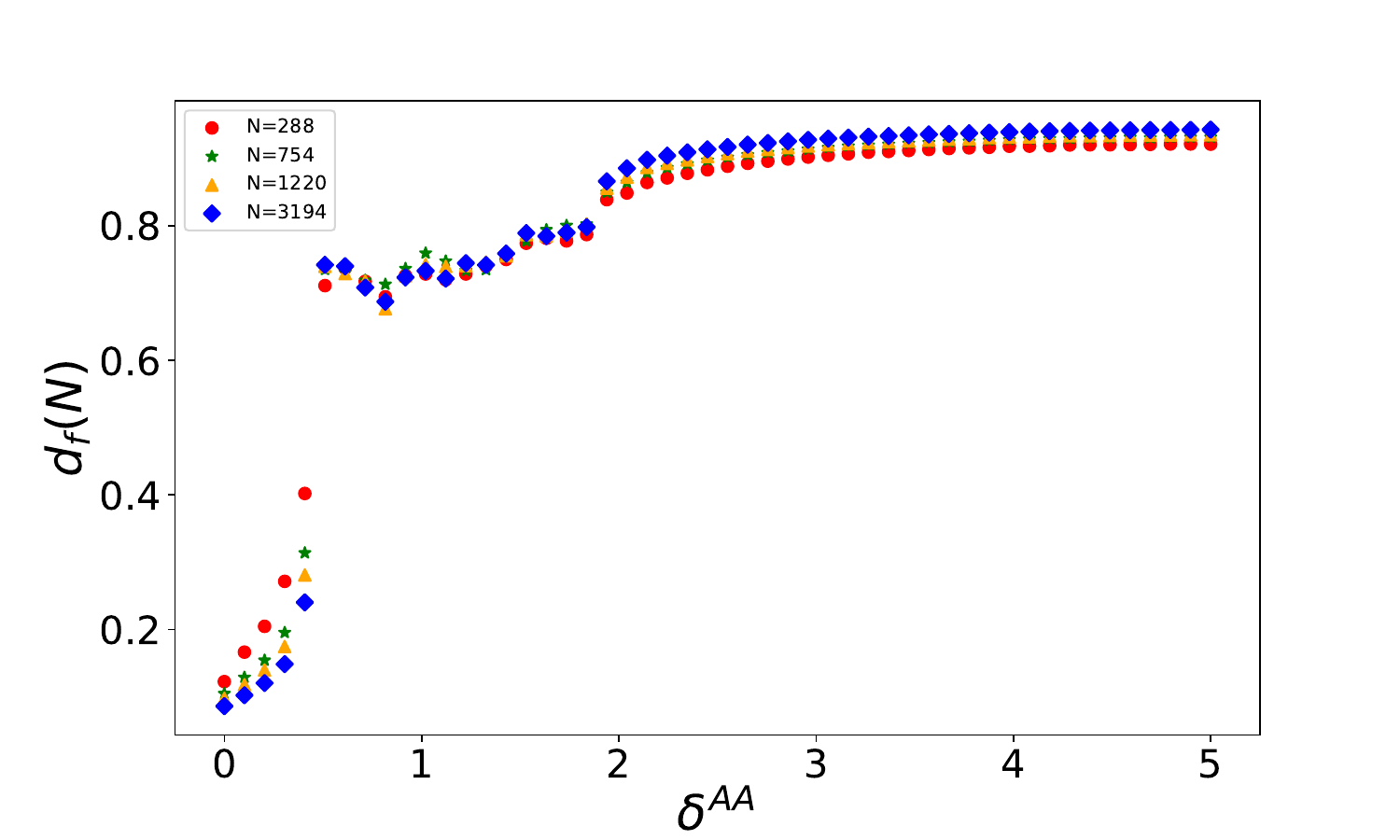}
    \includegraphics[keepaspectratio, width=0.45\textwidth]{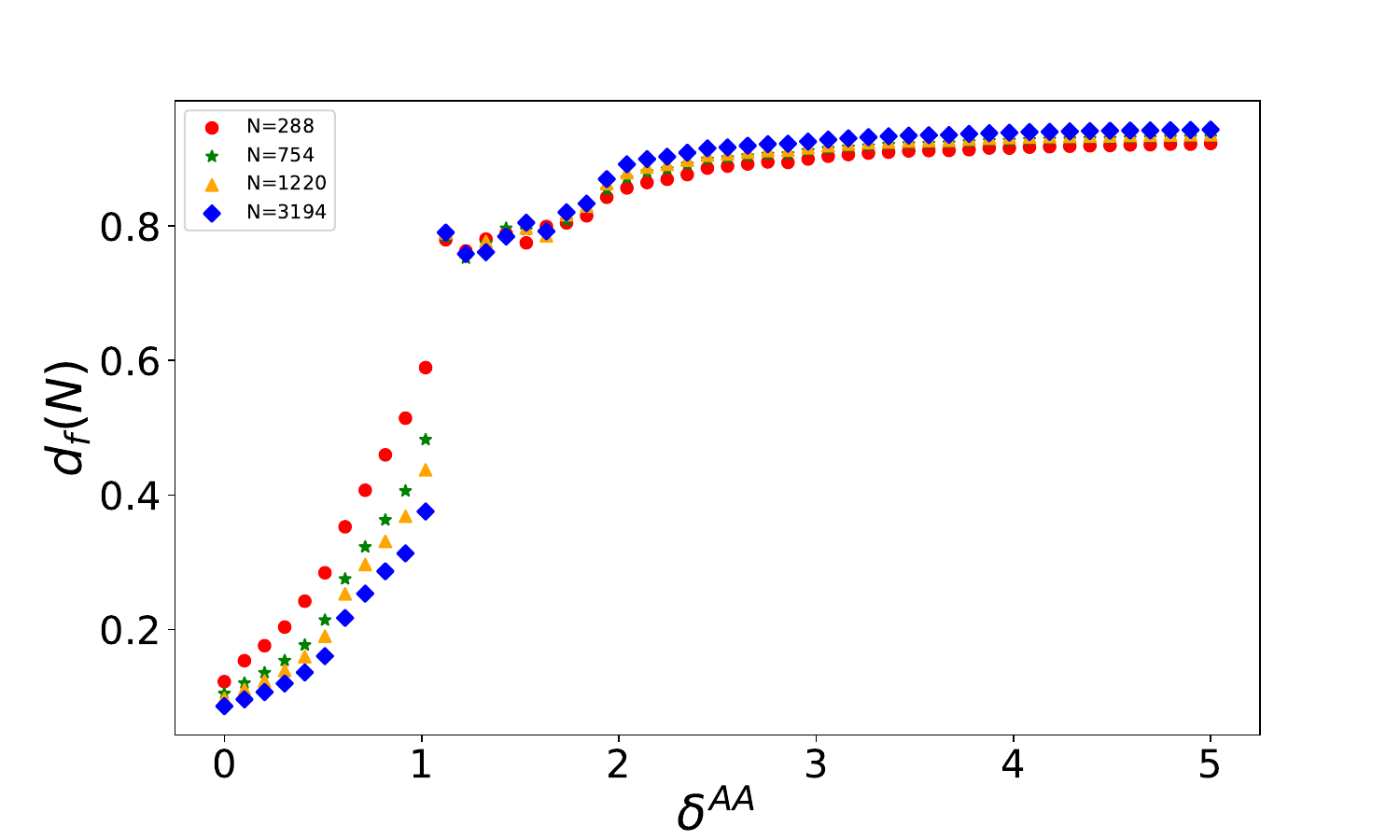}
    \put(-360,80){\textbf{(c) }}
    \put(-130,80){\textbf{(d) }}
    \\
    \caption{Plot of fractal dimension as a function of $\delta^{AA}$ for $\phi=0$ for various system sizes. (a): $K_{b}>K_{a}$ (real space), (b): $K_{a}>K_{b}$ (real space), (c): $K_{b}>K_{a}$ (momentum space), (d): $K_{a}>K_{b}$ (momentum space). In cases (a) and (b) the eigenstates of the system starts from being in an extended phase and make transition very fast into a localized phase as $\delta^{AA}$ increases. In cases (c) and (d), the eigenstates of the system starts from a localized phase and make transitions in a slow manner into an extended phase. In case (c), the kind of critical states (where the fractal dimension $d_{f}(N)$ does not converge to either $0$ or $1$) exist well some below $\delta^{AA}<1$ and extends for some $\delta^{AA}>2$, while in case (d) the value of $d_{f}(N)$ converge neither to $0$ or $1$ in the parameter range $\delta^{AA}>1$ and $\delta^{AA}<2$, which well matches with the parameter range of the TAI phase.}
    \label{fig:combined_vector_plot3}
\end{figure*}

Similarly for type $B$ mass in $j$th unit cell will be 
\begin{equation}
    \begin{split}
   \ddot{u}^{B}_{j} &= -[K_{a} + K_{b} + \delta^{AA} \cos(2\pi \beta (j) + \phi)+K^{0,B}_{j}] u^{B}_{j} \\
   &\quad + [K_{a} + \delta^{AA} \cos(2\pi \beta j + \phi)] u^{A}_{j} + [K_{b}] u^{A}_{j+1},
\end{split}
\end{equation}
where $K^{0,A}_{j}=\delta^{AA}[1-\cos(2\pi \beta j+\phi)]=K^{0,B}_{j}$ .
This selection of on-site spring stiffness preserves the chirality of the matrix and guarantees that all eigenvalues are positive. Assuming a normal mode solution, the two equations can be expressed as follows:
\begin{equation}
\begin{aligned}
\omega^{2}u^{A}_{j} &= (K_{a} + K_{b} + \delta^{AA})u^{A}_{j} \\
                     &\quad -(K_{a} +\delta^{AA} \cos(2\pi \beta (j) + \phi) )u^{B}_{j} \\
                     &\quad -(K_{b})u^{B}_{j-1} ,
\end{aligned}
\end{equation}
\vspace{0.25cm}   similarly for $B$ type mass one has 
\begin{equation}
\begin{aligned}
\omega^{2}u^{B}_{j} &= (K_{a} + K_{b} + \delta^{AA} )u^{B}_{j} \\
                     &\quad -(K_{a} + \delta^{AA} \cos(2\pi \beta (j) + \phi))u^{A}_{j} \\
                     &\quad -(K_{b})u^{A}_{j+1}.
\end{aligned}
\end{equation}
The two equations can be compactly expressed in matrix form using a dynamical matrix, similar to Eq.~\eqref{eq:dynamicalmat1} as
\begin{widetext}
{\small
\begin{equation}\label{eq:dynamicalmat2}
  D=\begin{bmatrix}
K_{a}+K_{b}+\delta^{AA} & -K_{a}-\delta^{AA}\cos(2\pi\beta+\phi) & \dots & -K_{b} \\
-K_{a}-\delta^{AA}\cos(2\pi\beta+\phi) & K_{a}+K_{b}+\delta^{AA} & -K_{b} & \dots \\
\vdots & \vdots & \ddots & -K_{a}-\delta^{AA}\cos(2\pi\beta(N)+\phi) \\
-K_{b} & \dots & -K_{a}-\delta^{AA}\cos(2\pi\beta(N)+\phi) & K_{a}+K_{b}+\delta^{AA} \\
\end{bmatrix}  . 
\end{equation}
}
\end{widetext}

\begin{figure*}[htbp]
    \centering
    \includegraphics[keepaspectratio, width=0.45\textwidth]{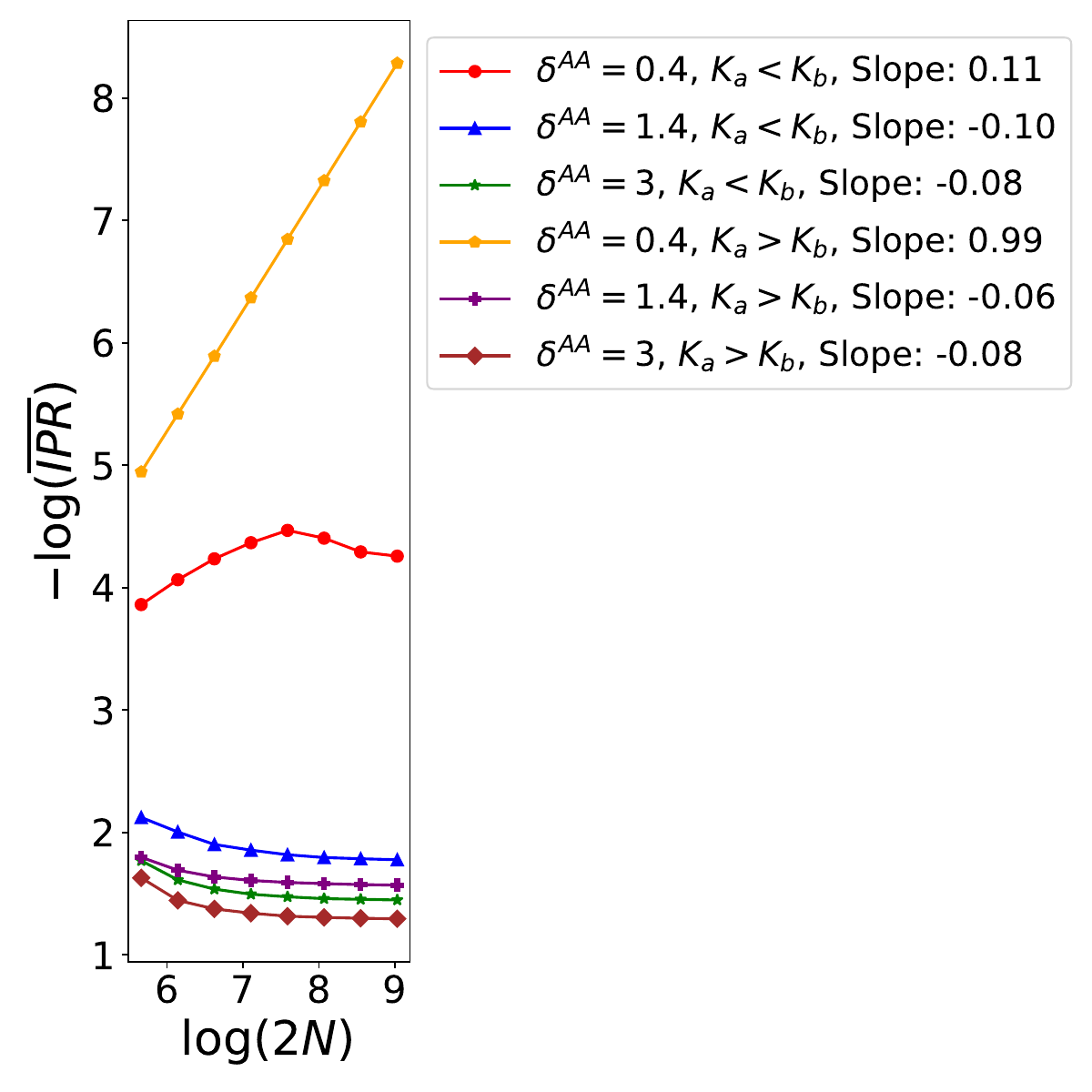}
    \includegraphics[keepaspectratio, width=0.45\textwidth]{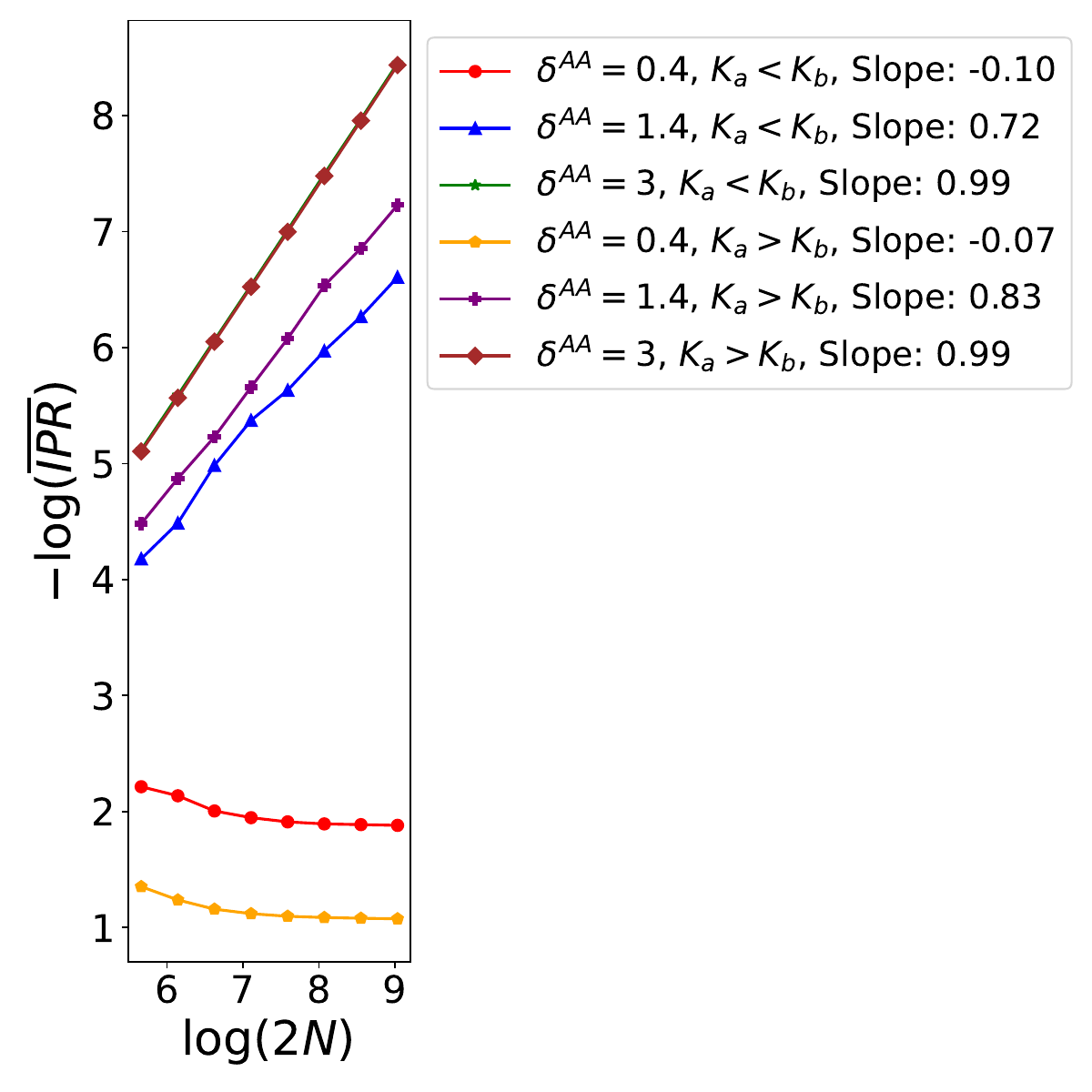}
     \put(-400,100){\textbf{(a) }}
     \put(-165,100){\textbf{(b)}}
     \\
     \caption{Plot of $-\ln(\overline{IPR})$ versus $\ln(2N)$ for $\phi=0$ in the cases of (a) real space and (b) momentum space. The slope of the curve indicates the value of $d_{f}(N)$, which as $N$ increases, converges to the fractal dimension. In case (b), the violet and blue curves indicate that the eigenstates of the system are in a critical phase, while violet curve is particularly interesting due to the TAI phase and provides a quantitative estimation of the presence of critical states in the TAI phase.}
    \label{fig:combined_vector_plot4}
\end{figure*}
Conductance for this model is computed using the non-equilibrium Green's function technique, as illustrated in Fig.~\ref{fig:e2B}.
In order to achieve genuine quasi-periodicity, it is important to work with an infinite system. To simulate this, we apply periodic boundary conditions within the lattice and find a suitable rational approximation $\tilde{\beta}=\frac{F_{n-1}}{F_{n}}$, where the numerator and denominator are successive terms in the Fibonacci sequence.
\subsection{Topological characterisation }\label{sec:b}
We characterize the topology by computing the LTM, as indicated in Eq.~\eqref{eq:LTM1}, for each $K_{a}$ and $\delta^{AA}$ while fixing $K_{b}=1$. This approach leads to convergence with the winding number. Specifically, we set $\phi=0$, use the rational approximation $\tilde{\beta}=\frac{377}{610}$, include $N=610$ unit cells, and vary $K_{a}\in[0,5]$ and $\delta^{AA}\in[0,6]$. The topological phase transition in clean systems i.e. with $\delta^{AA}=0$ occurs when $K_{a}=K_{b}$ \cite{asboth2016short}. Secondly, we will examine the behavior of localization length, which is known to diverge at the Fermi level during topological phase transitions in electronic systems \cite{PhysRevLett.113.046802,LIU2022128004,PhysRevLett.112.206602}. This divergence aligns with the closing of the band gap. To validate this, we analytically calculate the localization length $\Lambda$ and verify its divergence in the transition region. We measure the entire eigenvalue spectrum relative to the middle frequency $\omega_{0}^{2}=K_{a}+K_{b}+\delta^{AA}$, so that the Fermi level can be centered at zero energy level i.e. $\omega^{2}-\omega_{0}^{2}=0$. Hence the equation of motions for states at the Fermi level are:
\begin{equation}
    (K_{a}+\delta^{AA}\cos(2\pi\beta j+\phi))u^{B}_{j}+(K_{b})u^{B}_{j-1}=0,
\end{equation}
\begin{widetext}
    \begin{equation}
    (K_{a}+\delta^{AA}\cos(2\pi\beta j+\phi))u^{A}_{j}+(K_{b} + \delta^{AA}\cos(2\pi \beta (j) + \phi))u^{A}_{j+1}=0.
\end{equation}
\end{widetext}

Simplifying the equation for $A$ sub-lattice we have
\begin{equation}
    u^{A}_{j+1}=\left(\frac{-(K_{a}+\delta^{AA}\cos(2\pi\beta j+\phi))}{K_{b}}\right)u^{A}_{j}.
\end{equation}
The above equation leads to recursive relation as 
\begin{equation}\label{eq:19}
    u^{A}_{N}=(-1)^{N-1}\prod_{j=1}^{j=N-1}\left(\frac{-(K_{a}+\delta^{AA}\cos(2\pi\beta j+\phi))}{K_{b}}\right)u^{A}_{1}.
\end{equation}
We can derive the Lyapunov exponent $\gamma$, which is the inverse of the localization length \cite{PhysRevLett.113.046802,SCALES199727},
\begin{equation}
    \gamma=-\lim\limits_{N \to \infty}\frac{1}{N}\ln\left|\frac{u^{A}_{N}}{u^{A}_{1}}\right|,
\end{equation}
\begin{equation}\label{eq:lyapunon1}
    \gamma=-\lim\limits_{N \to \infty}\frac{1}{N}\sum_{j=1}^{N-1}\ln \left|\frac{K_{b}}{K_{a}+\delta^{AA}\cos(2\pi\beta j+\phi)}\right|.
\end{equation}
Due to the AA modulation being incommensurate with the underlying lattice, the summation can be expressed as a Riemann integral. Utilizing the properties of irrational rotation \cite{06bacaea-d5a8-3177-a6ba-b77181df9af3}, integrating it over angular variable $\theta \in [0, 2\pi]$, one has
\begin{equation}\label{eq:integral1}
    \gamma=-\frac{1}{2\pi}\int_{0}^{2\pi}\ln\left|\frac{K_{b}}{K_{a}+\delta^{AA}\cos(\theta)}\right|d\theta .
\end{equation}
The Eq.~\eqref{eq:integral1} can be evaluated to 
\begin{equation}\label{eq:Bound1}
    \gamma=\ln \left(\frac{K_{a}+\sqrt{K_{a}^{2}-(\delta^{AA})^{2}}}{2K_{b}}\right) ,
\end{equation}

for $K_{a}>\delta^{AA}$.
And for $K_{a}<\delta^{AA}$ the expression evaluate to 
\begin{equation}\label{eq:Bound2}
    \gamma=\ln\left(\frac{\delta^{AA}}{2K_{b}}\right) .
\end{equation}
Setting the Lyapunov exponent to $0$ computes the boundary of the topological phase transition, indicating a divergent localization length.
Equations for the critical diverging line are 
\begin{equation}
    K_{a}=\frac{(\delta^{AA})^{2}}{4K_{b}}+K_{b},
\end{equation} 

for $K_{a}>\delta^{AA}$ 
and 
\begin{equation}
    \delta^{AA}=2K_{b}
\end{equation} for $K_{a}<\delta^{AA}$.
The initial phase $\phi$ in the modulation does not need to vary, as its information is lost during integration with incommensurate modulation. The localization length at the central frequency can be numerically computed using the transfer matrix method \cite{SCALES199727} as a function of relevant parameters.

\subsection{ Calculation of inverse participation ratio (IPR)}
To gain insight into the localization properties of the model, we can analyze the inverse-participation ratios ($IPR$). The $IPR$ for a specific eigenstate is defined as 

\begin{equation}
    IPR(\psi\rangle)=\sum_{j=1}^{2N}|u_{j}|^{4},
\end{equation}
where \(u_{j}\) are the components of the normalized eigenstate \(|\psi\rangle\) in a space of dimension \(2N\).
For extended eigenmodes, the inverse participation ratio (IPR) scales as $\frac{1}{N}$, where $N$ is the system size. In contrast, fully localized eigenmodes exhibit an IPR that scales as $1$. Eigenmodes that do not fit these patterns define the critical phase of the phase transition \cite{RevModPhys.80.1355,PhysRevLett.126.106803,PhysRevB.105.174206,PhysRevB.109.014210}. The $IPR$ of eigenstates can be plotted against energy eigenvalues and Aubry-André strength $\delta^{AA}$. These plots facilitate the visualization of transitions among non-localized, localized, and critical phases in both real and momentum spaces.
When studying localization with $AA$ modulation, a crucial parameter to calculate is fractal dimension $D_{f}$, which does not depend on the size of the system, and influences the behavior of $IPR$ (or average of all state $IPR$) as size $N$ varies. The scaling relation is $\overline{IPR} \approx (2N)^{-D_{f}}$. To derive $D_{f}$ from finite size systems, we use the size-dependent quantity $d_{f}(N)$, which approaches the fractal dimension $D_{f}$ as $N$ increases. The expression for $d_{f}(N)$ is as follows:
\begin{equation}
    d_{f}(N)=-\frac{\ln(\overline{IPR})}{\ln(2N)}.
\end{equation}
In the context where $2N$ denotes the system size, $D_{f}$ equals $1$ in the non-localized phase and $0$ in the localized phase \cite{lu2022exact,PhysRevB.109.195427,sircar2024disorderdriventopologicalphase}. Numerical computations as shown in 
\begin{figure*}[htbp]
    \centering
    \includegraphics[keepaspectratio, width=0.45\textwidth]{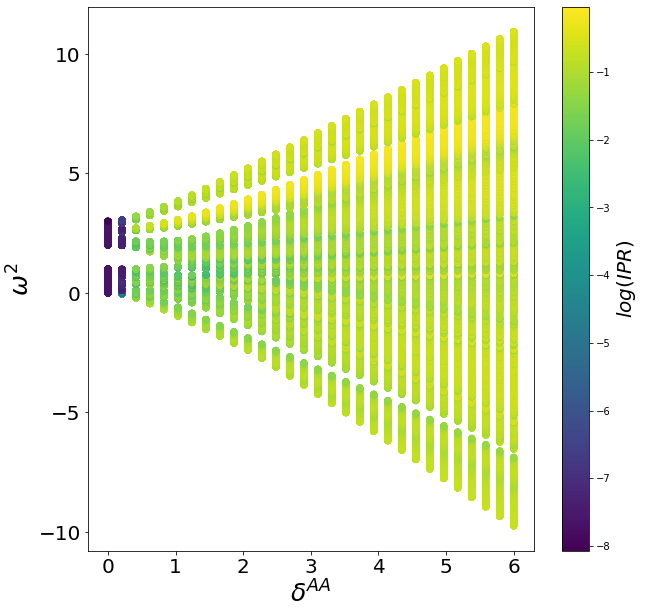}
    \includegraphics[keepaspectratio, width=0.45\textwidth]{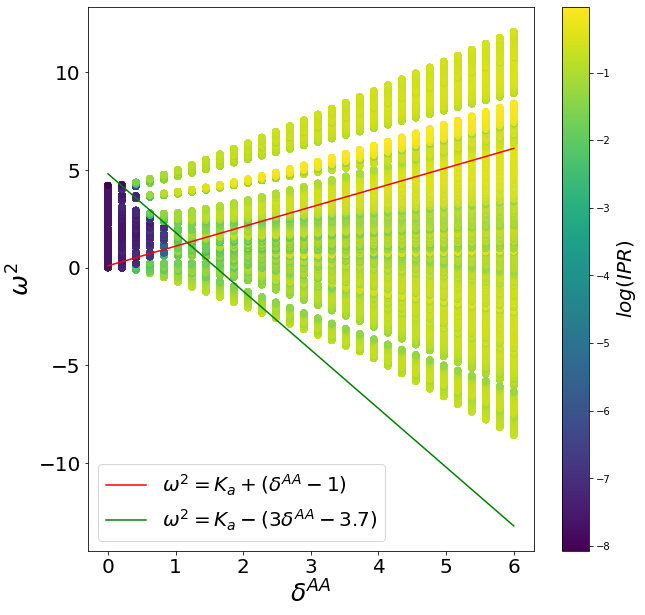}
     \put(-400,100){\textbf{(a) }}
     \put(-165,100){\textbf{(b)}}
     \\
     \caption{Plot of $\ln(IPR)$ versus the eigenvalue spectrum $\omega^{2}$ for $N=1597$ unit cells and $\phi=0$ in the original non-chiral model, to capture the complete eigenvalue dependence for subsequent cases: (a) $K_{a}<K_{b}$, (b) $K_{a}>K_{b}$. In case (b) one can observe the presence of critical states bounded by mobility edges with in the TAI phase regime, also shown in Fig.~\ref{fig:combined_vector_plot2}. We try to fit the energy dependence of mobility edge using trial and error method. The fitting curve can be expressed as $\omega^{2}=K_{a}+(\delta^{AA}-\alpha)$ for the lower mobility edge and $\omega^{2}=K_{a}-(\beta\delta^{AA}-\eta)$ for the upper mobility edge. Through trial and error, one can find that the values of $\alpha=1$, $\beta=3$, and $\eta=3.7$ match well with the mobility edges. These expressions are exact for a slowly varying quasi-periodic modulation. This article does not provide an exact expression for the mobility edge in the case of on-site Aubry-André quasi-periodic modulation. }
    \label{fig:combined_vector_plot7}
\end{figure*}
Fig.~\ref{fig:combined_vector_plot4}, shows that as $N$ increases, the value of $d_{f}(N)$ approaches $D_{f}$. The value of $d_{f}(N)$ converges to $1$ in case of the non-localized phase and to $0$ in case of the localized phase, where a finite portion of states remains localized. In the critical region, $d_{f}(N)$ stabilizes at value between $0$ and $1$.

\section{CONCLUSION}
In this study, we expanded the exploration of topological phase transitions in a mechanical SSH model, as described in \cite{sircar2024disorderdriventopologicalphase}, by applying Aubry-André modulation only to the intra-cellular spring constants, rather than the intercellular ones. This approach resulted in distinct localization properties and the emergence of mobility edges, which could not be analytically derived from Avila's global theory \cite{10.1007/s11511-015-0128-7}. We have derived the precise analytic expression for the anomalous mobility edge under inter-cellular Aubry-André quasi-periodic modulation in the article \cite{sircar2024disorderdriventopologicalphase}. We preserved the chiral symmetry of the dynamical matrix throughout our calculations. Consequently, we utilized a topological invariant from class-AIII or DBI, along with its real-space covariant form, to analyze and characterize the topological phases. We developed an analytical expression for the Lyapunov exponent, enabling us to predict the boundary of the topological phase transition based on system parameters ($K_{a}$, $K_{b}$, $\delta^{AA}$). We examined the localization properties of the eigenmodes by calculating the inverse participation ratio ($IPR$) and fractal dimensions. We have used fractal dimension \cite{lu2022exact,PhysRevB.109.195427,sircar2024disorderdriventopologicalphase} to quantitatively estimate the localized and non-localized phases as well as to detect the presence of critical phase. \par
The article emphasized that the global theory \cite{10.1007/s11511-015-0128-7} did not enable the exact calculation of mobility edges in such a scenario. We planned to develop novel analytical techniques for calculating the exact expression of mobility edge in our future work. \par
\section{Acknowledgement}
The author acknowledge support of the Department of
Atomic Energy, Government of India, under
Project Identification No. RTI4007. The author also thank Dr. Kabir Ramola for reviewing the article.

\bibliographystyle{apsrev4-2}
\bibliography{Bibliography4}

\appendix 
\section{Eigenvalues and eigenvectors for a symmetric circulant matrix}
\subsection{Periodic boundary condition}\label{sec:2}
Given the matrix with periodic boundary condition (PBC) as :
\begin{equation}
    D_{1}=\begin{pmatrix}
        2k & -k & 0 & \cdots & -k \\
        -k & 2k & -k & \cdots & 0 \\
        0 & -k & 2k & \ddots & \vdots \\
        \vdots & \vdots & \ddots & \ddots & -k \\
        -k & 0 & \cdots & -k & 2k
    \end{pmatrix}.
\end{equation}
The matrix structure reveals that all elements in each row and column are $0$, and that all rows are cyclic permutations of the first row.
So, the eigenvectors should be of the form
\begin{equation}
    |\psi_{m}\rangle=\begin{pmatrix}
        1\\
        x^{m}\\
        x^{2m}\\
        \vdots\\
        x^{(N-1)m}
    \end{pmatrix},
\end{equation}
where $x=e^{\frac{2\pi i}{N}}$.
Expanding the characteristic equation i.e $D_{1}|\psi_{m}\rangle=\lambda_{m}|\psi_{m}\rangle$ in each row, all the bulk equations i.e $i\in[2,N-1]$, are equivalent, where $i$ denotes the row number. The first and last row are equivalent only if $x^{N}=1$, which implies $x=e^{\frac{2\pi i}{N}}$.
Hence from the expansion of last row :
\begin{equation}
    \lambda_{m}=2-x^{m}-x^{m(N-1)},
\end{equation}
which determines the $m$th eigenvalue as 
\begin{equation}
    \lambda_{m}=2-e^{\frac{2\pi i m}{N}}-e^{\frac{2\pi i m(N-1)}{N}}=2(1-\cos(\frac{2\pi m}{N})).
\end{equation}
\begin{figure}[ht!]
    \centering
    \includegraphics[width=\linewidth]{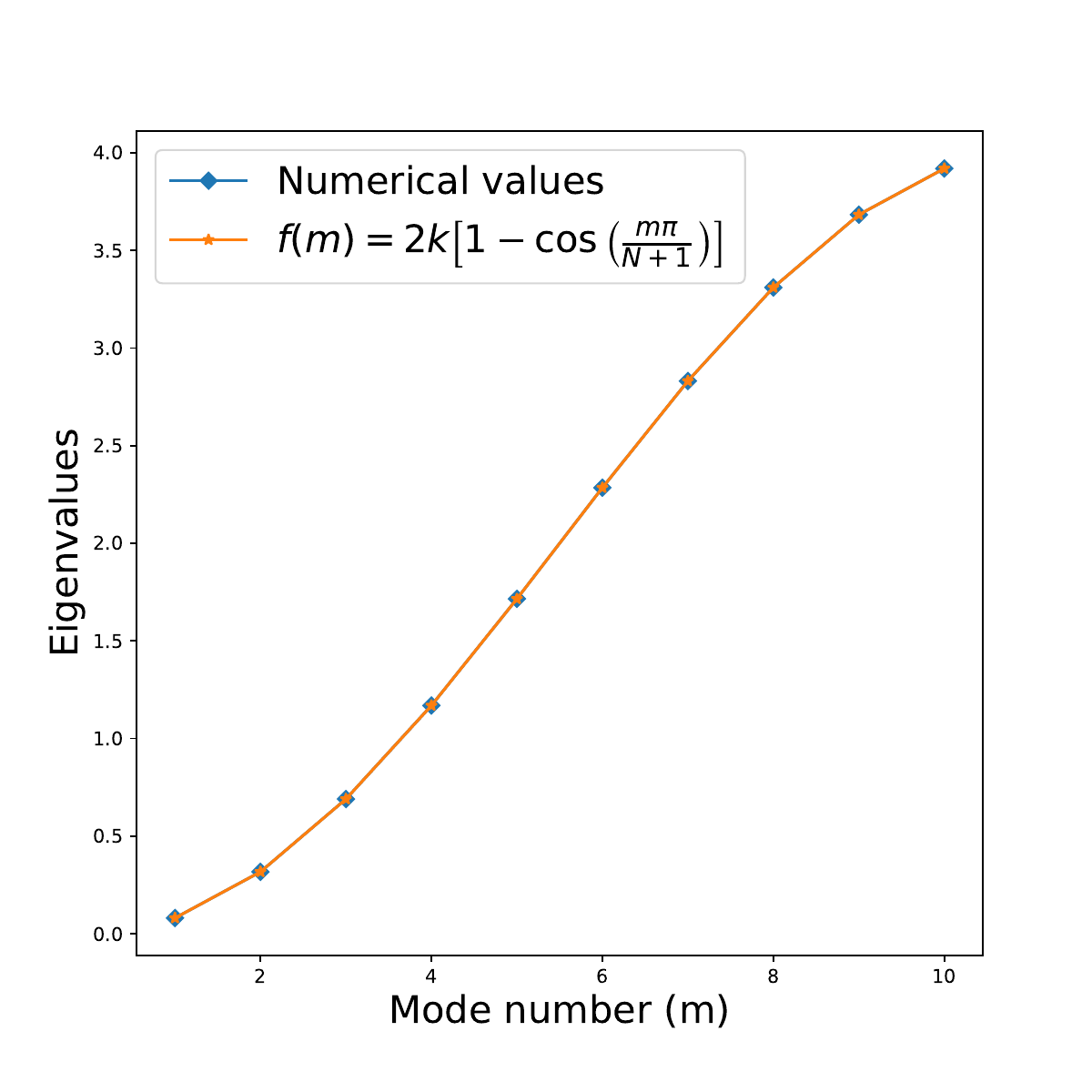}
    \caption{Plot depicting that the eigenvalues computed numerically for a symmetric circulant matrix of size $10\times10$ with fixed boundary condition matches with the analytical results obtained in Eq.~\eqref{eq:9}. The $\delta$ scale indicates the difference between the numerical and analytical computation.}
    \label{fig:1}
\end{figure}
\begin{figure}[ht!]
    \centering
    \includegraphics[width=\linewidth]{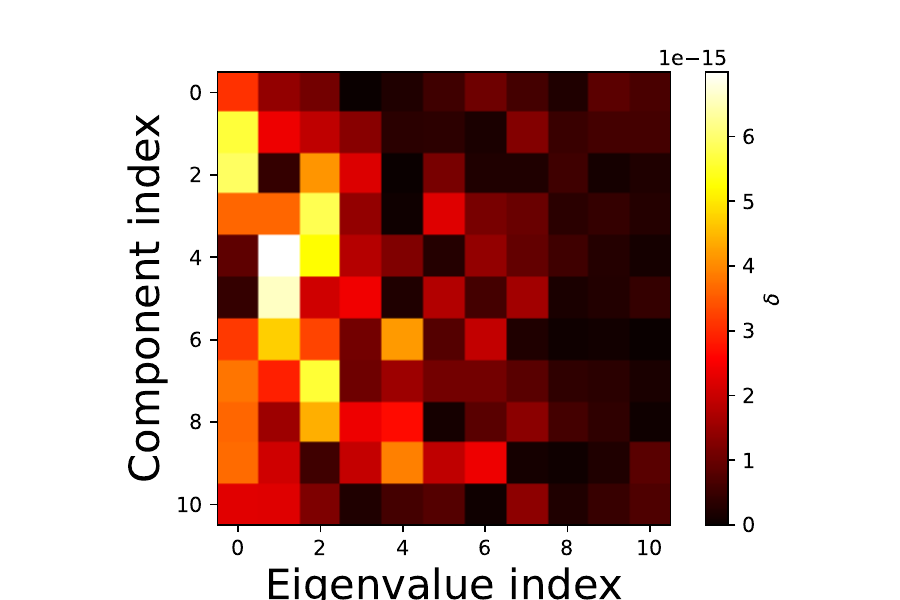}
    \caption{Plot depicting that the eigenvectors computed numerically for a symmetric circulant matrix of size $10\times10$ with fixed boundary condition matches with the analytical results obtained in Eq.~\eqref{eq:10}.The $\delta$ scale indicates the difference between the numerical and analytical computation.}
    \label{fig:2}
\end{figure}
\subsection{Fixed boundary conditions }
With fixed boundary conditions (FBC) the matrix elements get modify as given below:
\begin{equation}
    D_{2}=\begin{pmatrix}
        2k & -k & 0 & \cdots & 0 \\
        -k & 2k & -k & \cdots & 0 \\
        0 & -k & 2k & \ddots & \vdots \\
        \vdots & \vdots & \ddots & \ddots & -k \\
        0 & 0 & \cdots & -k & 2k
    \end{pmatrix}.
\end{equation}
By substituting $t=\frac{2k-\lambda}{k}$, the matrix simplifies to the following form:
\begin{equation}
    D_{2}=\begin{pmatrix}
        t & -1 & 0 & \cdots & 0 \\
        -1 & t & -1 & \cdots & 0 \\
        0 & -1 & t & \ddots & \vdots \\
        \vdots & \vdots & \ddots & \ddots & -1 \\
        0 & 0 & \cdots & -1 & t
    \end{pmatrix}.
\end{equation}
Denoting the determinant of this $N\times N$ as $\Delta_{N}$, one can write a recurssion relation by expanding the determinant as :
\begin{equation}\label{eq:Det1}
    \Delta_{N}=\det\begin{vmatrix}
        t & -1 & 0 & \cdots & 0 \\
        -1 & t & -1 & \cdots & 0 \\
        0 & -1 & t & \ddots & \vdots \\
        \vdots & \vdots & \ddots & \ddots & -1 \\
        0 & 0 & \cdots & -1 & t
    \end{vmatrix},
\end{equation}
\begin{equation}
    \Delta_{N}=t\Delta_{N-1}-\Delta_{N-2},
\end{equation}
with initial condition given as $\Delta_{0}=1$, $\Delta_{1}=t$, obtain from Eq.~\eqref{eq:Det1}.
The function $\Delta_{n}(t)$ follows the same recursive relation and initial conditions as the $n$th order Chebyshev polynomial of the first kind. Thus, the $m$th eigenvalue can be expressed in terms of the $m$th zeros of the Chebyshev polynomial of the first kind as:
\begin{equation}
    t_{m}=2\cos(\frac{m\pi}{N+1}),
\end{equation}
so, the eigenvalues are given as 
\begin{equation}\label{eq:9}
    \lambda_{m}=2k-t_{m}k=2k(1-\cos(\frac{m\pi}{N+1})),
\end{equation}
for $m\in[1,N]$.
Now, let the $m$th eigenvector be written as $|\psi\rangle=(x_{1},x_{2},x_{3},\cdots,x_{N})^{T}$, then the equations for the components are :
\begin{equation}
    (2k-\lambda_{m})x_{1}+kx_{2}=0,
\end{equation}
which implies $x_{2}=\frac{\lambda_{m}-2k}{k}=-T_{1}(\lambda_{m})$, taking $x_{1}=1$.
From the second row one can obtain 
\begin{equation}
    kx_{1}+(2k-\lambda_{m})x_{2}+kx_{3}=0,
\end{equation}
which implies $x_{3}=\frac{\lambda_{m}^{2}-4k^{2}}{k^{2}}-1=T_{2}(\lambda_{m})=(-1)^{2}T_{2}(\lambda_{m})$.
Hence continuing in this manner, using mathematical induction, it is easy to verify that the eigenvectors can be written as :
\begin{equation}\label{eq:10}
    |\psi_{m}\rangle=\begin{pmatrix}
        T_{0}(\lambda_{m})\\
        (-1)T_{1}(\lambda_{m})\\
        (-1)^{2}T_{2}(\lambda_{m})\\
        \vdots\\
        (-1)^{N-1}T_{N-1}(\lambda_{m})
    \end{pmatrix}
\end{equation}
\section{Eigenvalues and eigenvectors for the mechanical SSH dynamical matrix with fixed boundary conditions}\label{sec:appendixB}
The matrix representation with fixed boundary condition is 
\begin{equation}
    D_{2k}=\begin{pmatrix}
        K_{a}+k_{b} & -K_{a} & 0 & \cdots & 0 \\
        -K_{a} & K_{a}+K_{b} & -K_{b} & \cdots & 0 \\
        0 & -K_{b} & K_{a}+K_{b} & \ddots & \vdots \\
        \vdots & \vdots & \ddots & \ddots & -K_{a} \\
        0 & 0 & \cdots & -K_{a} & K_{a}+K_{b}
    \end{pmatrix}.
\end{equation}
Let $a_{1} = -K_{a}$, $a_{2} = -K_{b}$, and $u = K_{a} + K_{b}$. Then, the determinant can be expressed as a recursion relation:
\begin{equation}\label{eq:Reec1}
    |D_{2k}-\lambda I|=(u-\lambda)|D_{2k-1}-\lambda I|-a_{1}^{2}|D_{2k-2}-\lambda I|,
\end{equation} and 
\begin{equation}\label{eq:Reec2}
    |D_{2k+1}-\lambda I|=(u-\lambda)|D_{2k}-\lambda I|-a_{2}^{2}|D_{2k-1}-\lambda I|,
\end{equation}
Starting from the case of $n=1$, we have simplified the determinants to :
\begin{equation}\label{eq:1}
    |D_{2k}-\lambda I|=a_{1}^{2k}f_{k}(t),
\end{equation} and 
\begin{equation}\label{eq:2}
    |D_{2k+1}-\lambda I|=a_{1}^{2k}(u-\lambda)g_{k}(t),
\end{equation}
where $t=\frac{u-\lambda}{a_{1}^{2}}$.
Then the recursion relation can be written as :
\begin{equation}
    a_{1}^{2k}f_{k}(t)=(u-\lambda)^{2}a_{1}^{2(k-1)}g_{k-1}(t)-a_{1}^{2}a_{1}^{2(k-1)}f_{k-1}(t),
\end{equation}
which can be simplified as :
\begin{equation}
    a_{1}^{2k}f_{k}(t)=\frac{(u-\lambda)^{2}}{a_{1}^{2}}a_{1}^{2k}g_{k-1}(t)-a_{1}^{2k}f_{k-1}(t),
\end{equation}
\begin{equation}\label{eq:4}
    f_{k}(t)=tg_{k-1}(t)-f_{k-1}(t).
\end{equation}
Similarly from Eq.~\eqref{eq:Reec2} and using Eq.~\eqref{eq:1} with Eq.~\eqref{eq:2} we can write :
\begin{equation}
    a_{1}^{2k}(u-\lambda)g_{k}(t)=(u-\lambda)a_{1}^{2k}f_{k}(t)-a_{2}^{2}a_{1}^{2(k-1)}(u-\lambda)g_{k-1}(t),
\end{equation}
\begin{equation}
    g_{k}(t)=f_{k}(t)-\frac{a_{2}^{2}}{a_{1}^{2}}g_{k-1}(t)=tg_{k-1}(t)-f_{k-1}(t)-\frac{a_{2}^{2}}{a_{1}^{2}}g_{k-1}(t),
\end{equation}
\begin{equation}\label{eq:3}
    g_{k}(t)=(t-\frac{a_{2}^{2}}{a_{1}^{2}})g_{k-1}(t)-f_{k-1}(t)=(t-v)g_{k-1}(t)-f_{k-1}(t),
\end{equation} where
$v=\frac{a_{2}^{2}}{a_{1}^{2}}$.
From Eq.~\eqref{eq:3} by simplifying we can obtain:
\begin{equation}\label{eq:15}
    f_{k-1}(t)=(t-v)g_{k-1}(t)-g_{k}(t),
\end{equation}
\begin{equation}
    g_{k+1}(t)=(t-v)g_{k}(t)-f_{k}(t).
\end{equation}
Now putting those in Eq.~\eqref{eq:4}:
\begin{equation}
    f_{k}(t)=(t-v)g_{k}(t)-g_{k+1}(t),
\end{equation}
\begin{equation}
    (t-v)g_{k}(t)-g_{k+1}(t)=tg_{k-1}(t)-(t-v)g_{k-1}(t)-g_{k}(t),
\end{equation}
\begin{equation}
    (t-v+1)g_{k}(t)-g_{k+1}(t)=vg_{k-1}(t),
\end{equation}
which leads to the following recursion relation:
\begin{equation}\label{eq:5}
    g_{k+1}(t)=(t-v+1)g_{k}(t)-vg_{k-1}(t).
\end{equation}
From Eq.~\eqref{eq:4} one can write by substituting $k+1$ in place of $k$, one can write:
\begin{equation}
    tg_{k}(t)=f_{k+1}(t)+f_{k}(t).
\end{equation}
Similarly using Eq.~\eqref{eq:3}:
\begin{equation}
    tg_{k}(t)=(t-v)tg_{k-1}(t)-tf_{k-1}(t),
\end{equation}
\begin{equation}
    tg_{k}(t)=(t-v)(f_{k}(t)+f_{k-1}(t))-tf_{k-1}(t),
\end{equation}
\begin{equation}
    tg_{k}(t)=(t-v)f_{k}(t)+f_{k-1}(t)(t-v-t),
\end{equation}
\begin{equation}
    f_{k+1}(t)+f_{k}(t)=(t-v)f_{k}(t)-vf_{k-1}(t),
\end{equation}
which leads to following similar recursion relation for function $f(t)$ as:
\begin{equation}\label{eq:6}
    f_{k+1}(t)=(t-v-1)f_{k}(t)-vg_{k-1}(t).
\end{equation}
From Eq.~\eqref{eq:6} by rearranging terms one can write:
\begin{equation}\label{eq:7}
    \frac{1}{\sqrt{v}}f_{k+1}(t)=\frac{t-v-1}{\sqrt{v}}f_{k}(t)-\sqrt{v}f_{k-1}(t),
\end{equation}
By substituting, 
\begin{equation}
    F_{k}(w)=\frac{1}{v^{\frac{k}{2}}}f_{k}(t),
\end{equation} where
$w=\frac{t-v-1}{\sqrt{v}}=\frac{t-l^{2}-1}{l}$,
\begin{equation}
    F_{k}(w)=\frac{1}{l^{k}}f_{k}(t)=\frac{1}{l^{k}}f_{k}(1+l^{2}+wl).
\end{equation}
Hence from Eq.~\eqref{eq:7} one can obtain,
\begin{equation}
    \frac{1}{\sqrt{v}}v^{\frac{k+1}{2}}F_{k+1}(w)=wv^{\frac{k}{2}}F_{k}(w)-v^{\frac{k-1}{2}}\sqrt{v}F_{k-1}(w),
\end{equation}
which leads to the following recursion relation for $F(w)$ as :
\begin{equation}
    F_{k+1}(w)=wF_{k}(w)-F_{k-1}(w),
\end{equation} 
with initial conditions $F_{0}(w)=1$ and $F_{1}(w)=w+l$.
A similar recursion relation can be obtained for function $G_{k}(w)=\frac{1}{v^{\frac{k}{2}}}g_{k}(t)$ as :
\begin{equation}
    G_{k+1}(w)=wG_{k}(w)-G_{k-1}(w),
\end{equation}
with initial conditions $G_{0}(w)=1$ and $G_{1}(w)=w$.
Hence, from above we can see that with same initial conditions the function $G_{k}(w)$ is a Chebyshev polynomial of order $k$ of first kind , whose zeroes are given as 
\begin{equation}
    w_{m}=2\cos(\frac{m\pi}{k+1}),
\end{equation} where 
$m\in[1,k]$.
Hence the expression for $t$ will be :
\begin{equation}
    t_{m}=1+2l\cos(\frac{m\pi}{k+1})+l^{2}.
\end{equation}
So the eigenvalues for $N=2k+1$ is given as :
\begin{equation}
    (K_{a}+K_{b}-\lambda_{m})=+(-)\sqrt{K_{a}^{2}+K_{b}^{2}+2K_{a}K_{b}\cos(\frac{m\pi}{k+1})},
\end{equation}
which simplifies to 
\begin{equation}\label{eq:8}
    \lambda_{m}=(K_{a}+K_{b})-(+)\sqrt{K_{a}^{2}+K_{b}^{2}+2K_{a}K_{b}\cos(\frac{m\pi}{k+1})}.
\end{equation}
which is similar to the dispersion relation of original tight-binding SSH model with two different hopping parameters $K_{a}$ and $K_{b}$ respectively.
Similarly for $N=2k$ the eigenvalues are given as :
\begin{equation}
    \lambda_{m}=(K_{a}+K_{b})-(+)\sqrt{K_{a}^{2}+K_{b}^{2}+K_{a}K_{b}z_{m}},
\end{equation} where 
$z_{m}$ is determined using $F_{k+1}(w)=0$ for $w=z_{m}$.
Finding \( z_{m} \) is subtle. We aim to express the polynomial in terms of the Chebyshev polynomial \( G_{k}(w) \) to simplify the process of locating the zeroes \( z_{m} \).
From Eq.~\eqref{eq:1} 
\begin{equation}
    |D_{2k}-\lambda I|=a_{1}^{2k}f_{k}(t)=(K_{a})^{2k}((t-v-1)f_{k-1}(t)-vf_{k-2}(t)),
\end{equation}
From Eq.~\eqref{eq:15} similarly one can write,
\begin{equation}
    f_{k}(t)=(t-v)g_{k}(t)-g_{k+1}(t),
\end{equation}
\begin{equation}
    f_{k}(t)=(t-v)g_{k}(t)-(t-v+1)g_{k}(t)+vg_{k-1}(t),
\end{equation}
simplifying further,
\begin{equation}
    f_{k}(t)=-g_{k}(t)+vg_{k-1}(t).
\end{equation}
Therefore,
\begin{widetext}
    \begin{equation}
    |D_{2k}-\lambda I|=(K_{a})^{2k}(-g_{k}(t)+vg_{k-1}(t))=(K_{a}K_{b})^{k}(-G_{k}(w)+(\frac{K_{b}}{K_{a}})^{\frac{1}{2}}G_{k-1}(w)).
\end{equation}
\end{widetext}

We obtain the determinant expansion using Chebyshev polynomials $G_{k}(w)$. Setting this to $0$ yields: 
\begin{equation}
    G_{k}(w)=(\frac{K_{b}}{K_{a}})^{\frac{1}{2}}G_{k-1}(w),
\end{equation}
This allows for the determination of $z_{m}$, though an exact expression for it remains elusive.
The eigenvector can be obtained by considering term by term as :
\begin{equation}
    (K_{a}+K_{b}-\lambda_{m})x_{1}-K_{a}x_{2}=0,
\end{equation}
\begin{equation}
    x_{2}=\frac{(K_{a}+K_{b})-\lambda_{m}}{K_{a}}=\frac{(K_{a}+K_{b})-\lambda_{m}}{K_{a}}G_{0}(z_{m}),
\end{equation} with $x_{1}=1=F_{0}(z_{m})$.
Similarly from second row one can write:
\begin{equation}
    -K_{a}x_{1}+(K_{a}+K_{b}-\lambda_{m})x_{2}-K_{b}x_{3}=0,
\end{equation}
\begin{equation}
    x_{3}=\frac{(K_{a}+K_{b}-\lambda_{m})^{2}-K_{a}^{2}}{K_{a}K_{b}}x_{1}=F_{1}(z_{m})x_{1}.
\end{equation}
So continuing in this manner and using mathematical induction we have obtain the eigenvectors as :
\begin{equation}
    |\psi_{m}\rangle=\begin{pmatrix}
        F_{0}(z_{m})\\
        \frac{K_{a}+K_{b}-\lambda_{m}}{K_{a}}G_{0}(z_{m})\\
        F_{1}(z_{m})\\
        \frac{K_{a}+K_{b}-\lambda_{m}}{K_{a}}G_{1}(z_{m})\\
        \vdots\\
        \frac{K_{a}+K_{b}-\lambda_{m}}{K_{a}}G_{k-1}(z_{m})
    \end{pmatrix}.
\end{equation}
Similarly the eigenvector for eigenvalues given in Eq.~\eqref{eq:8} as :
\begin{equation}
    |\psi_{m}\rangle=\begin{pmatrix}
        F_{0}(w_{m})\\
        \frac{K_{a}+K_{b}-\lambda_{m}}{K_{a}}G_{0}(w_{m})\\
        F_{1}(w_{m})\\
        \frac{K_{a}+K_{b}-\lambda_{m}}{K_{a}}G_{1}(w_{m})\\
        \vdots\\
        \frac{K_{a}+K_{b}-\lambda_{m}}{K_{a}}G_{k-1}(w_{m})\\
        F_{k}(w_{m})
    \end{pmatrix}.
\end{equation}
Here $G_{k}$'s are nothing but Chebyshev polynomial $T_{k}$ as noted above.
Hence the eigenvectors can be written as :
\begin{equation}
    |\psi_{m}\rangle=\begin{pmatrix}
        F_{0}(z_{m})\\
        \frac{K_{a}+K_{b}-\lambda_{m}}{K_{a}}T_{0}(z_{m})\\
        F_{1}(z_{m})\\
        \frac{K_{a}+K_{b}-\lambda_{m}}{K_{a}}T_{1}(z_{m})\\
        \vdots\\
        \frac{K_{a}+K_{b}-\lambda_{m}}{K_{a}}T_{k-1}(z_{m})
    \end{pmatrix},
\end{equation} for $N=2k$,
\begin{equation}
    |\psi_{m}\rangle=\begin{pmatrix}
        F_{0}(w_{m})\\
        \frac{K_{a}+K_{b}-\lambda_{m}}{K_{a}}T_{0}(w_{m})\\
        F_{1}(w_{m})\\
        \frac{K_{a}+K_{b}-\lambda_{m}}{K_{a}}T_{1}(w_{m})\\
        \vdots\\
        \frac{K_{a}+K_{b}-\lambda_{m}}{K_{a}}T_{k-1}(w_{m})\\
        F_{k}(w_{m})
    \end{pmatrix},
\end{equation} for $N=2k+1$.

\end{document}